\documentclass[aps,tightenlines,showpacs]{revtex4}
\usepackage{graphicx}
\usepackage{float}
\def\lg{{\mathchoice{~\raise.58ex\hbox{$<$}\mkern-14.8mu\lower.52ex\hbox{$>$}~}
                    {~\raise.58ex\hbox{$<$}\mkern-14.8mu\lower.52ex\hbox{$>$}~}
                    {\raise.59ex\hbox{{$\scriptscriptstyle <$}}\mkern-12.8mu%
                     \lower.01ex\hbox{{$\scriptscriptstyle >$}}}   {}   }} 
\def\gl{{\mathchoice{~\raise.58ex\hbox{$>$}\mkern-12.8mu\lower.52ex\hbox{$<$}~}
                    {~\raise.58ex\hbox{$>$}\mkern-12.8mu\lower.52ex\hbox{$<$}~}
                    {\raise.62ex\hbox{{$\scriptscriptstyle >$}}\mkern-12.0mu%
                     \lower.05ex\hbox{{$\scriptscriptstyle <$}}}  {}    }}   

\def\ca{{\mathchoice{~\raise.58ex\hbox{$c$}\mkern-9.0mu\lower.52ex\hbox{$a$}~}
                    {~\raise.58ex\hbox{$c$}\mkern-9.0mu\lower.52ex\hbox{$a$}~}
                    {\raise.59ex\hbox{{$\scriptscriptstyle c$}}\mkern-7.0mu%
		     \lower.01ex\hbox{{$\scriptscriptstyle a$}}}   {} 	}} 
\def\ac{{\mathchoice{~\raise.58ex\hbox{$a$}\mkern-10.0mu\lower.52ex\hbox{$c$}~}
                    {~\raise.58ex\hbox{$a$}\mkern-10.0mu\lower.52ex\hbox{$c$}~}
		    {\raise.62ex\hbox{{$\scriptscriptstyle a$}}\mkern-9.0mu%
		     \lower.05ex\hbox{{$\scriptscriptstyle c$}}}  {} 	}}   

\newcommand{\be}{\begin{equation}}
\newcommand{\ee}{\end{equation}}
\newcommand{\ba}{\begin{eqnarray}}
\newcommand{\ea}{\end{eqnarray}}
\newcommand{\comma}{\;, \\ [2mm] \nonumber}

\newcommand{\per}{\;.}
\newcommand{\bold}[1]{\mbox{\boldmath $#1$}} 
 
\newcommand{\ra}{\rightarrow} 

\begin{document}

\title{Transport Theory beyond Binary Collisions}

\author{Margaret E. Carrington\footnote{Electronic address:
{\tt carrington@brandonu.ca}}}

\affiliation{Department of Physics, Brandon University, \\
Brandon, Manitoba, R7A 6A9 Canada\\
and Winnipeg Institute for Theoretical Physics,\\
Winnipeg, Manitoba, Canada}

\author{Stanis\l aw Mr\' owczy\' nski\footnote{Electronic address:
{\tt mrow@fuw.edu.pl}}}

\affiliation{So\l tan Institute for Nuclear Studies, \\
ul. Ho\.za 69, PL - 00-681 Warsaw, Poland \\
and Institute of Physics, \'Swi\c etokrzyska Academy, \\
ul. \'Swi\c etokrzyska 15, PL - 25-406 Kielce, Poland}

\date{January 10, 2005}
\begin{abstract}

Using the Schwinger-Keldysh technique, we derive the transport equations 
for a system of quantum scalar fields. We first discuss the general 
structure of the equations and then their collision terms. Taking into 
account up to three-loop diagrams in $\phi^3$ model and up to four-loop
diagrams in $\phi^4$ model, we obtain transport equations which include 
the contributions of multi-particle collisions and particle production 
processes, in addition to mean-field effects and binary interactions.

\end{abstract}


\maketitle

\section{Introduction}

Transport theory is a very convenient tool to study many-body 
nonequilibrium systems, both relativistic and nonrelativistic. 
The kinetic equations, which play a central role in the transport 
approach, usually assume that dissipation processes are governed 
by binary collisions. However, when the system of interest is very 
dense one expects that multi-particle interactions will play a 
significant role. Such interactions are known to the control spectra
of fluctuations and transport properties of dense gases and liquids
\cite{Han76}. Furthermore, in relativistic systems a characteristic 
particle's kinetic energy is usually comparable to the particle's mass, 
and processes leading to particle production become important for 
a system's dynamics. 

It is therefore expected that multi-particle interactions and production processes will play an important role in the dynamics of a relativistic 
quark-gluon plasma of high energy density. The importance of gluon 
multiplication in the process of the plasma's equilibration has been 
repeatedly stressed, see {\it e.g.} \cite{Xiong:1992cu,Geiger:1991nj,Srivastava:1998vj,Baier:2000sb,Arnold:2002zm,Heinz:2004ik}. The scattering of three gluons into three gluons has
been studied in the context of thermalization in \cite{Xu:2004gw}. 
It has also been shown within the scalar field theory that the transport
coefficient of bulk viscosity, as given by the Kubo formula, strongly 
depends on particle number changing processes \cite{Jeon:1994if,Jeon:1995zm}. 
We conclude that a complete description of the quark gluon plasma based 
on transport theory requires the derivation of the relativistic transport
equation which includes the multi-particle collisions and particle production 
processes, in addition to the mean-field effects and binary interactions.
The form of such an equation has been postulated by some authors, see 
{\it e.g.} \cite{Arnold:2002zm,Xu:2004gw,Jeon:1994if,Jeon:1995zm}, but 
a systematic derivation from first principles is lacking even in the 
simplest scalar field theory.

The Schwinger-Keldysh \cite{Schwinger:1960qe,Keldysh:ud} formulation of 
quantum field theory provides a framework for the derivation of the 
transport equation. Kadanoff and Baym \cite{Kad62} developed a technique 
for nonrelativistic quantum systems which has been further generalized 
to relativistic systems \cite{Bez72,Li:1982gk,Danielewicz:kk,Calzetta:1986cq,Mrowczynski:1989bu,Botermans:1990qi,Mrowczynski:1992hq,Henning:sm,Boyanovsky:1996xx,Boyanovsky:1999cy,Klevansky:1997wm}. 
We mention here only the papers which go beyond the mean field or 
Vlasov approximation and provide a more or less systematic analysis 
of the (binary) collision term. 

In this paper, we consider self-interacting scalar fields with cubic 
and quartic interaction terms and study their transport equations beyond 
the binary collision approximation. There are basically two separate parts 
of our study. The first part is the derivation of the general structure of 
the transport equations. Such an analysis has been published previously by 
one of us \cite{Mrowczynski:1989bu,Mrowczynski:1992hq,Mrowczynski:1997hy} 
and it is presented here only to provide a framework for the second part 
of the paper. The main steps of the derivation are the following. We define 
the contour Green's function with the time arguments on the contour in 
the complex time plane. This function is a key element of the Schwinger-Keldysh 
approach. After discussing its properties and relevance for nonequilibrium 
systems, we write down the exact equations of motion {\it i.e.} the 
Dyson-Schwinger equations. We perform a Wigner transformation and do a 
gradient expansion by assuming that the system has macroscopic 
quasi-homogeneity. The resulting pair of Dyson-Schwinger equations is 
converted into the transport and mass-shell equations which are satisfied 
by the Wigner function. We define the distribution function of usual 
probabilistic interpretation, and we find the transport equations satisfied 
by this function. The transport equation is derived to lowest order in 
the gradient expansion, which corresponds to the Markovian limit in which 
memory terms are neglected.

In the second part of our study, we perform a perturbative analysis of 
the self-energies which enter the transport equation. We show that the Vlasov 
or mean-field terms are dominated by the lowest order tadpole diagrams, 
while binary collisions emerge from two-loop contributions. Multi-particle 
interactions and production processes appear at three-loop level in the 
$\phi^3$ model and at four-loops in the $\phi^4$ model.  There are obviously
higher loop contributions to both mean field and binary collision terms 
but these higher loop contributions will not change these effects 
qualitatively, and for this reason are not studied here. Instead we obtain
leading order contributions to the collision term that result from 
$2 \leftrightarrow 3$ processes in $\phi^3$ model, and 
$2 \leftrightarrow 4$ and $3 \leftrightarrow 3$ processes in 
$\phi^4$ model. To extract the processes represented by the three- 
and four-loop graphs, we work with the Keldysh representation, 
using a technique developed previously by one of us 
\cite{Carrington:jt,Carrington:2002bv}.

Multi-particle and production processes have already been discussed in the 
Keldysh-Schwinger approach to transport theory but, to our best knowledge, 
only for nonrelativistic systems \cite{Bez68,Dan90,Bozek:1997rv}. In this case  
the analysis is quite different. In the nonrelativistic limit, 
interactions are instantaneous and internal lines representing the potential 
interaction cannot be cut. Consequently, the extraction of the scattering 
matrix elements in the small coupling limit is considerably simpler than 
in the relativistic case. However, it should be noted that much progress 
has been achieved in resuming multi-loop diagrams in the nonrelativistic 
approach \cite{Dan90}.

The problem of off-mass-shell transport has recently attracted a lot attention
in the literature \cite{Ivanov:1999tj,Ivanov:2003wa,Leupold:1999ga,Leupold:xe,Juchem:2003bi,Juchem:2004cs}. 
The collision term for off-mass-shell transport contains important 
contributions from one- and two-loop self-energy diagrams which correspond 
to $1 \leftrightarrow 2$ and $1 \leftrightarrow 3$ processes. 
These processes are kinematically forbidden for on-mass-shell particles, 
and for this reason are not studied here.

Throughout this article we use natural units where $\hbar = c = 1$. 
The signature of the metric tensor is $(+,-,-,-)$.

\section{Preliminaries}
\label{preliminaries}

We consider a system of real scalar fields with Lagrangian 
density of the form
\begin{equation}\label{lagran}
{\cal L}(x) = {1 \over 2}\partial^{\mu}\phi (x)\partial_{\mu}\phi (x)
- {1 \over 2}m^{2} \phi^{2}(x)
- {g \over n!}\phi^n(x) \;,
\end{equation}
where $n$ equals 3 or 4. The renormalization counterterms are omitted  
here. The field satisfies the equation of motion
\begin{equation}\label{motion}
\bigr[ \partial^{2} + m^{2} \bigl] \phi (x) = 
- {g \over (n-1)!} \phi^{n-1}(x) \;. 
\end{equation}
The energy-momentum tensor is defined as 
$$
T^{\mu \nu}(x) = \partial^{\mu}\phi (x)\partial^{\nu}\phi (x)
- g^{\mu \nu}{\cal L}(x) \;.
$$
Subtracting the total derivative 
$$
{1 \over 4} \partial^{\mu}\partial^{\nu} \phi^{2}(x) - 
g^{\mu \nu}{1 \over 4} \partial^{\sigma}\partial_{\sigma}
\phi^{2}(x) \;,
$$ 
and considering the case of free fields, we obtain an expression for 
the energy-momentum tensor which has a form that is convenient for 
our purposes
\begin{equation}\label{tensor}
T^{\mu \nu}_0 (x)= - {1 \over 4} \phi (x)\buildrel \leftrightarrow
\over \partial^{\mu} \buildrel \leftrightarrow
\over \partial^{\nu} \phi (x)  \;. 
\end{equation}

\section{Green's functions}
\label{G-functions}

\begin{figure}[H]
\begin{center}
\includegraphics[width=5cm]{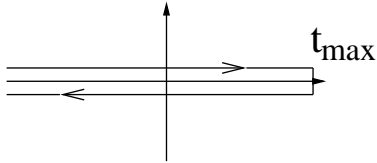}
\end{center}
\caption{\label{fig-contour} The contour in the complex time plane
for an evaluation of the operator expectation values.} 
\end{figure}

We define the contour Green's function as
\ba
\label{contour}
i\Delta (x,y) \buildrel \rm def \over 
= \langle  \tilde T \phi (x) \phi (y) \rangle \;, 
\ea
where $\langle ...\rangle $ denotes the ensemble average at time $t_0$ 
(usually identified with $-\infty $) and $\tilde T$ is the time ordering 
operation along the directed contour shown in Fig.~\ref{fig-contour}. 
The parameter $t_{\rm max}$ is shifted to $+ \infty$ in calculations. 
The time arguments are complex with an infinitesimal positive or negative 
imaginary part which locates them on the upper or lower branch of the 
contour. The ordering operation is defined as
\ba
\label{c-ordering}
\tilde T \phi (x) \phi (y) \buildrel \rm def \over = 
\Theta (x_0,y_0)\phi (x) \phi (y) +
\Theta (y_0,x_0)\phi (y) \phi (x) \;,
\ea
where $\Theta (x_0,y_0)$ equals 1 if $x_0$ succeeds $y_0$ on the 
contour, and  0 if $x_0$  precedes $y_0$. 

\par

If the field develops a finite expectation value, as is the case 
when the symmetry is spontaneously broken, the contribution 
$\langle \phi (x)\rangle \langle \phi (y) \rangle$ is subtracted 
from the right-hand-side of the equation defining the Green's function 
and one looks at field fluctuations around the expectation values 
(see {\it e.g.} \cite{Mrowczynski:1989bu,Mrowczynski:1992hq}). Since 
$\langle \phi (x)\rangle$ is expected to vanish in the models defined 
by the Lagrangians (\ref{lagran}) we neglect this contribution in the 
definition of the  Green's function. 

The contour propagator involves four functions, which can be thought of 
as corresponding to propagation  along the top branch of the contour, 
from the top branch to the bottom branch, from the bottom branch to the 
top branch, and along the bottom branch.  Below, we define four Green's 
functions with real time arguments, and discuss their relationship to 
the contour Green's function. We define
\ba \label{contour-parts}
i\Delta^> (x,y) &\buildrel \rm def \over =&
 \langle  \phi (x) \phi (y) \rangle \;, \\ [2mm] \nonumber 
i\Delta^<  (x,y) &\buildrel \rm def \over =&
\langle  \phi (y) \phi (x) \rangle \;, \\ [2mm] \nonumber 
i\Delta^c (x,y) &\buildrel \rm def \over =& 
\langle  T^c \phi (x) \phi (y) \rangle \;, \\ [2mm] \nonumber 
i\Delta^a (x,y) &\buildrel \rm def \over = &
\langle  T^a \phi (x) \phi (y) \rangle \;, 
\ea
where $T^c (T^a)$ indicates chronological (anti-chronological) time ordering
$$
T^c \phi (x) \phi (y) \buildrel \rm def \over = 
\Theta (x_0-y_0) \phi (x) \phi (y) +
\Theta (y_0-x_0) \phi (y) \phi (x) \;,
$$
$$
T^a \phi (x) \phi (y) \buildrel \rm def \over = 
\Theta (y_0-x_0) \phi (x) \phi (y) +
\Theta (x_0-y_0) \phi (y) \phi (x) \;.
$$
These functions are related to the contour Green's functions 
in the following manner:
\ba \label{ca><}
\Delta^c(x,y) \equiv \Delta (x,y) \;\; {\rm for} \;\;
&x_0& , \; y_0 \;\; {\rm on \;\; the \;\; upper \;\; branch,} 
\\ [2mm] \nonumber
\Delta^a(x,y) \equiv \Delta (x,y) \;\; {\rm for} \;\;
&x_0& , \; y_0 \;\; {\rm on \;\; the \;\; lower \;\; branch,} 
\\ [2mm] \nonumber 
\Delta^<(x,y) \equiv \Delta (x,y) \;\; {\rm for} \;\;
&x_0&  \; {\rm on \;\; the \;\; upper \;\; branch\;\; and \;\;} 
\\ \nonumber
&y_0& \;\; {\rm on \;\; the \;\; lower \;\; one,} 
\\ [2mm] \nonumber 
\Delta^>(x,y) \equiv \Delta (x,y) \;\; {\rm for} \;\;
&x_0& \;\; {\rm on \;\; the \;\; lower \;\; branch  \;\; and \;\;} 
\\ \nonumber
&y_0& \;\; {\rm on \;\; the \;\; upper \;\; one. } 
\ea
The top branch of the contour is usually called the `1' branch and the 
bottom branch is called the `2' branch. Using this notation,  
the functions $\Delta^c$, $\Delta^a$, $\Delta^<$ and $\Delta^>$ are
often written as
\ba
\label{1/2}
\Delta^c &=& \Delta_{11} \;, \;\;\;\;\; \Delta^a = \Delta_{22} \;,
\\ \nonumber
\Delta^< &=& \Delta_{12} \;, \;\;\;\;\, \Delta^> = \Delta_{21} \;.
\ea
The following identities can be obtained directly from the definitions given 
above:
\ba\label{ident-c}
\Delta^{\ca}(x,y) &=& \Theta (x_0 - y_0) \Delta^{\gl}(x,y) +
\Theta (y_0 - x_0) \Delta^{\lg}(x,y) \;, \\ [2mm] \nonumber
(i \Delta^{\ca}(x,y))^\dagger &=& i\Delta^{\ac}(x,y)
= i\Delta^{\ac}(y,x) \;, 
\\ [2mm] \nonumber
(i \Delta^{\gl}(x,y))^\dagger &=& i\Delta^{\gl}(x,y) 
= i\Delta^{\lg}(y,x) \;,
\ea
where the $\dagger$ denotes hermitian conjugation which involves 
an interchange of the arguments of the Green's function. Using 
Eqs.~(\ref{contour-parts}) and (\ref{ident-c}), it is easy to show that the 
four components of the contour Green's function are not independent but 
satisfy the relation:
\ba
\label{circling}
\Delta^c(x,y) + \Delta^a(x,y) -\Delta^<(x,y)-\Delta^>(x,y)=0
\ea

In some situations, it is useful to work with retarded $(+)$, advanced
$(-)$ and symmetric Green's functions. We make the usual definitions for 
these propagators:
\ba
\label{ret-xspace}
i\Delta^+(x,y) &\buildrel \rm def \over=& 
\Theta(x_0-y_0) \langle [\phi(x),\phi(y)] \rangle
= i\Theta(x_0-y_0) \big( \Delta^>(x,y) - \Delta^<(x,y) \big) \;, \\[2mm] 
\label{adv-xspace}
i\Delta^-(x,y) &\buildrel \rm def \over=& 
-\Theta(y_0-x_0) \langle [\phi(x),\phi(y)] \rangle
= i\Theta(y_0-x_0) \big( \Delta^<(x,y) - \Delta^>(x,y) \big) \;,
\\ [2mm] \label{sym-xspace}
i\Delta^{\rm sym}(x,y) &\buildrel \rm def \over=& 
\langle \{ \phi(x),\phi(y) \}\rangle
= i\big( \Delta^>(x,y) + \Delta^<(x,y) \big) \;,
\ea
where the curly brackets indicate an anticommutator. 

It is easy to show that there is a simple relationship between the 
original set of propagators ($\Delta^c$, $\Delta^a$, $\Delta^<$ and 
$\Delta^>$) and the retarded, advanced and symmetric propagators 
($\Delta^+$, $\Delta^-$ and $\Delta^{\rm sym}$). In fact, since the four 
original propagators satisfy the constraint (\ref{circling}) and
thus only three of them are independent, transferring between the two 
representations is equivalent to a change of basis. Henceforth, we will 
refer to the set of propagators $\Delta^+$, $\Delta^-$ and $\Delta^{\rm sym}$ 
as the Keldysh representation of the contour ordered propagator, and the set 
$\Delta^c$, $\Delta^a$, $\Delta^<$ and $\Delta^>$ as the 1/2 representation. 
Using the definitions (\ref{contour-parts}) and the relations (\ref{ident-c}), 
it is straightforward to connect the Keldysh to the 1/2 basis  
\ba
\label{ra-><}
\Delta^{\pm}(x,y) &=& \Delta^c(x,y)-\Delta^{\lg}(x,y) 
\comma
\Delta^{\gl}(x,y) &=& \frac{1}{2} 
\Big( \Delta^{\rm sym}(x,y) \pm \Delta^+(x,y) \mp \Delta^-(x,y) \Big) 
\comma
\Delta^+(x,y)&-&\Delta^-(x,y) = \Delta^>(x,y)-\Delta^<(x,y) \per
\ea

The relations (\ref{ident-c},\ref{circling},\ref{ra-><}) hold for real 
and complex fields. For pure real fields, which are studied here, there 
are extra relations among Green's functions. In particular, we will make 
use of the idendity 
\ba
\label{>vs<}
\Delta^>(x,y) = \Delta^<(y,x) \;.
\ea
 
We discuss briefly the physical interpretation of the Green's functions 
we have defined above. The functions $\Delta^{\ca}$ and $\Delta^{\pm}$
describe the propagation of a disturbance in which a single particle or 
antiparticle is added to the many-particle system at space-time point 
$y$ and then is removed from it at a space-time  point $x$. The function 
$\Delta^c(x,y)$ describes a particle disturbance propagating forward in 
time, and an antiparticle disturbance propagating backward in time. The 
meaning of $\Delta^a(x,y)$ is analogous but particles are propagated 
backward in time and antiparticles forward. In the zero density limit 
$\Delta^c(x,y)$ coincides with the Feynman propagator. In the case of 
the retarded (advanced) Green functions, both particles and antiparticles 
evolve forward (backward) in time. 

The physical meaning of functions $\Delta^\lg(x,y)$ is more transparent
when one considers their Wigner transforms which are given by
\begin{equation}\label{Wigner}
\Delta^{\lg}(X,p) \buildrel \rm def \over = \int d^4u \; e^{ipu}
\Delta^{\lg}(X+{1 \over 2}u,X-{1 \over 2}u) \;.
\end{equation}
It is easy to show that the free-field energy-momentum tensor 
(\ref{tensor}) averaged over the ensemble can be expressed as
\begin{equation}\label{en-mom-free}
\langle T^{\mu \nu}_0(X) \rangle = 
\int {d^4p \over (2\pi)^{4}} p^{\mu} p^{\nu} i\Delta^{\lg}(X,p) \;.
\end{equation}
One recognizes the standard form of the energy-momentum tensor in 
kinetic theory with the function $i\Delta^{\lg}(X,p)$ playing the role of  
the density of particles with four-momentum $p$ at the space-time point $X$. 
This observation leads us to consider $i\Delta^{\lg}(X,p)$ as the quantum 
analog of the classical distribution function. The function 
$i\Delta^{\lg}(X,p)$ is hermitian but it is not positive definite, and 
thus the probabilistic interpretation is only approximately valid. One 
should also note that, in contrast to the classical distribution function, 
the function $i\Delta^{\lg}(X,p)$ can be nonzero for the off-mass-shell 
four-momenta. As will be explicitly shown in Sec. \ref{distri-f}, one 
can extract the usual distribution function from $i\Delta^{\lg}(X,p)$ 
under specific conditions.

\section{Exact Equations of Motion}
\label{eq-of-motion}

Since the Green's function $i\Delta^{\lg}(X,p)$ will be used to obtain
the distribution function, we derive an equation which governs the 
evolution of $i\Delta^{\lg}(X,p)$. We start with the Dyson-Schwinger 
equation for the {\em contour} Green's function which has the form
\ba
\label{DS}
\Delta(x,y) = \Delta_0(x,y) -
\int_C d^4z \int_C d^4z' \Delta_0(x,z)\Pi(z,z')\Delta(z',y) \;,
\ea
where $\Delta_0(x,y)$ is the free Green's function. We note that the 
integrals over $z_0$ and $z_0'$ in the last term of equation (\ref{DS}) 
are performed along the contour. 

The Dyson-Schwinger equation (\ref{DS}) can be rewritten as two equations:
\begin{equation}\label{DS1}
\big[\partial_x^{2} + m^{2}\big] \Delta (x,y) = 
-\delta^{(4)}(x,y) + \int_C d^4x' \Pi(x,x')
\Delta(x',y) \;, 
\end{equation}
\begin{equation}\label{DS2}
\big[\partial_y^{2} + m^{2}\big] \Delta (x,y) = 
-\delta^{(4)}(x,y) + \int_C d^4x' \Delta(x,x')
\Pi(x',y) \;,
\end{equation}
where the function 
$\delta^{(4)}(x,y)$ is defined on the contour as 
\begin{displaymath}
\delta^{(4)}(x,y) = \left\{ \begin{array}{ccl} 
\delta^{(4)}(x-y) \;\;\; & {\rm for} &\;\; x_0 \;, \; y_0 \;\; 
{\rm from \;\; the \;\; upper \;\; branch,} \\
0 \;\;\;\;\;\;\; & {\rm for} &\;\; x_0 \;, \; y_0 \;\; 
{\rm from \;\; the \;\; different \;\; branches,} \\ 
-\delta^{(4)}(x-y) \;\;\; & {\rm for} & \;\; x_0 \;,\; y_0 \;\; 
{\rm from \;\; the \;\; lower \;\; branch.} \end{array} \right. 
\end{displaymath}

To obtain equations for $\Delta^{\lg}$ we split the self-energy 
into three parts as 
\be \label{self-split}
\Pi (x,y) = \Pi_{\delta}(x)\delta^{(4)}(x,y) + \Pi^>(x,y) \Theta (x_0,y_0)
+ \Pi^<(x,y) \Theta (y_0,x_0) \;, 
\ee
where $\Pi_{\delta}$ corresponds to the tadpole contribution to the 
self-energy. With the help of the retarded and advanced Green's functions
(\ref{ret-xspace},\ref{adv-xspace}) and similar expressions for the 
retarded and advanced self-energies, we can rewrite equations 
(\ref{DS1},\ref{DS2}) to obtain
\begin{eqnarray}\label{DSgl1}
\big[ \partial^{2}_x + m^{2} - \Pi_{\delta}(x)\big ] \Delta^{\gl}(x,y) 
&=& \int d^4x' \Bigr[ \Pi^{\gl }(x,x') \Delta^{-}(x',y)+
\Pi^{+}(x,x') \Delta^{\gl}(x',y) \Bigl]  \;, 
\\[2mm] \label{DSgl2}
\big[ \partial^{2}_y  + m^{2} - \Pi_{\delta}(y) \big] 
\Delta^{\gl }(x,y) 
&=&\int d^4x' \Bigr[ \Delta^{\gl}(x,x') \Pi^{-}(x',y)+
\Delta^{+}(x,x') \Pi^{\gl}(x',y)\Bigl]  \;, 
\end{eqnarray}
where all time integrations run from $- \infty$ to $+ \infty$.

Similarly, the Dyson-Schwinger equations (\ref{DS1},\ref{DS2}) provide 
the equations satisfied by the retarded and advanced functions 
$\Delta^{\pm}(x,y)$ as
\begin{eqnarray}\label{DSpm1}
\big[\partial^{2}_x + m^{2} - \Pi_{\delta}(x)\big] 
\Delta^{\pm}(x,y) &=& -\delta^{(4)}(x-y)+
\int d^4x'  \Pi^{\pm}(x,x') \Delta^{\pm}(x',y) \;,
\\ [2mm] \label{DSpm2}
\big[\partial^{2}_y + m^{2} - \Pi_{\delta}(y)\big] 
\Delta^{\pm}(x,y)  
&=& -\delta^{(4)}(x-y) + 
\int d^4x'  \Delta^{\pm}(x,x') \Pi^{\pm}(x',y) \;. 
\end{eqnarray}
We note that the equations for the retarded and for advanced functions
are decoupled from each other - there is no mixing of the retarded, 
advanced and symmetric components of the propagators. 

\section{Approximations}
\label{approx}

It is a very difficult task to study systems which are strongly
inhomogeneous or strongly interacting - there are no general methods 
applicable to such systems. Thus, we assume that systems of interest 
are weakly inhomogeneous and weakly interacting. We discuss below 
the mathematical formulation of the corresponding approximations.

\begin{enumerate}

\item[{\bf 1)}] The system is assumed to be weakly inhomogeneous 
in comparison with two different length scales.

\begin{enumerate}

\item We assume that the inhomogeneity length is large 
compared to the inverse of the characteristic momentum {\it i.e.}
\ba
\label{gradient-approx}
|F(X,p)| \gg |{\partial^{2} \over \partial X_{\mu} \partial p_{\nu}}
F(X,p)|  \;,
\ea
where $F$ is either the propagator or the self-energy. This assumption
allows one to perform the gradient expansion. 

\item The inhomogeneity length of the system is assumed 
to be large compared to the inverse characteristic mass of free 
quasi-particles (or the Comptom wavelength):
\ba
\label{quasi-particle-approx}
|F(X,p)| \gg {1 \over m^{2}} 
|{\partial^{2} \over \partial X_{\mu} \partial X^{\mu}}
F(X,p)|  \;.
\ea
This condition justifies the so-called quasi-particle approximation.
When the bare fields are massless or the free mass is much smaller 
than the dynamically generated effective mass $m_*$, the mass in
Eq.~(\ref{quasi-particle-approx}) should be replaced by $m_*$ 
\cite{Mrowczynski:1997hy}.

\end{enumerate}

\item[{\bf 2)}] The assumption that the system is weakly interacting 
means that the coupling constant $g$ in Eq.~(\ref{lagran}) is small 
and that all self-energies can be expanded perturbatively in $g$.

\end{enumerate}

The conditions of weak inhomogeneity and weak interaction will 
be used in five separate ways:

\begin{enumerate}

\item[{\bf (i)}] We will convert the equations (\ref{DSgl1}, \ref{DSgl2}) 
into transport equations by performing Wigner transformations (\ref{Wigner}) 
on all Green's functions and self-energies. Using Eq.~(\ref{gradient-approx}), 
we obtain the following set of translation rules: 
\begin{eqnarray}
\label{trans-rules}
\int d^4x' f(x,x') g(x', y) & \longrightarrow &
f(X,p)g(X,p)  +  {i \over 2} \{f(X,p),\,g(X,p)\} \;, \\
h(x) g(x, y) & \longrightarrow &
h(X)g(X,p) -  {i \over 2} {\partial h(X) \over \partial X^{\mu}}
{\partial g(X,p) \over \partial p_{\mu}}\;, \nonumber\\
h(y) g(x, y) & \longrightarrow &
h(X)g(X,p) +  {i \over 2} {\partial h(X) \over \partial X^{\mu}}
{\partial g(X,p) \over \partial p_{\mu}}\;,  \nonumber\\
\partial^{\mu}_x f(x,y) & \longrightarrow &
(-ip^{\mu} + {1 \over 2} \partial^{\mu})f(X,p) \;, \nonumber\\
\partial^{\mu}_y f(x,y) & \longrightarrow &
(ip^{\mu} + {1 \over 2} \partial^{\mu})f(X,p) \;, \nonumber
\end{eqnarray}
where we have introduced the Poisson-like bracket defined as
$$
\Big\{ C(X,p),\, D(X,p) \Big\} \equiv
{\partial C(X,p) \over \partial p_{\mu}}
{\partial D(X,p) \over \partial X^{\mu}} - 
{\partial C(X,p) \over \partial X^{\mu}}
{\partial D(X,p) \over \partial p_{\mu}} \;.
$$
The function $h(x)$ is weakly dependent on $x$ and we use the notation  
$\;\partial^{\mu} \equiv {\partial \over \partial X_{\mu}}$. This calculation 
is done in Sec. \ref{eqs-><}. 

\item[{\bf (ii)}]  The condition (\ref{quasi-particle-approx}) allows one
to drop the terms containing $\partial^2$. These terms and those 
involving ${\rm Im}\Pi^+$, which are discussed in (iv), are responsible 
for off-mass-shell contributions to the Green's functions $\Delta^{\gl}$.

\item[{\bf (iii)}] The assumption (\ref{gradient-approx}) is used together 
with the condition of weak interaction to show that the gradient terms in 
the right hand side of the transport equation (\ref{trans1}) and mass-shell 
equation (\ref{mass1}) can be dropped. This point is discussed in detail in 
Sec. \ref{eqs-><}.

\item[{\bf (iv)}] The weakly interacting approximation is used in 
Sec. \ref{mass-shell-con}. For $1/g \gg 1$ and a finite bare mass, we have  
\ba
\label{quasi-particle-approx-2}
m^2 - \Pi_\delta - {\rm Re}\Pi^+ \gg |{\rm Im}\Pi^+| \;.
\ea
This assumption combined with the condition (\ref{quasi-particle-approx})
allows us to obtain an expression for the spectral function that is 
proportional to  $\delta (p^2 - m_*^2)$. We note that both of equations 
(\ref{quasi-particle-approx}) and (\ref{quasi-particle-approx-2}) are 
commonly called `the quasi-particle approximation'. This terminology is 
a consequence of the fact that there are two independent means by which the 
quasi-particle picture can be destroyed: strong inhomogeneity and strong 
interactions.

\item[{\bf (v)}] The weakly interacting condition is also used in Sec. 
\ref{collision-term} to perform a perturbative expansion of the 
self-energies in the collision term of the transport equation. Each 
term in this expansion represents a different physical contribution 
to the transport equation.

\end{enumerate}

\section{Equations for $\Delta^{\gl}$}
\label{eqs-><}
 
Applying the translation rules (\ref{trans-rules}) to 
Eqs.~(\ref{DSgl1}, \ref{DSgl2}), neglecting the terms proportional 
to $\partial^{2}$ due to the quasi-particle approximation 
(\ref{quasi-particle-approx}), and taking the difference and sum 
of the equations, we obtain  
\begin{eqnarray}\label{trans}
\Bigr[p^{\mu} \partial_{\mu} - 
{1 \over 2} \partial_{\mu} \Pi_{\delta}(X) \partial^{\mu}_p \Bigl] 
\Delta^{\gl }(X,p)  
& = & {i \over 2} \Big( \Pi^>(X,p) \Delta^<(X,p) -
      \Pi^< (X,p) \Delta^> (X,p) \Big) \nonumber \\
& - & {1 \over 4} 
\Big\{ \Pi^{\gl}(X,p), \Delta^+(X,p) + \Delta^-(X,p) \Big\}  \nonumber \\
& - &  {1 \over 4} 
\Big\{ \Pi^+(X,p) + \Pi^-(X,p),\, \Delta^{\gl}(X,p) \Big\} \;,
\end{eqnarray}
\begin{eqnarray}\label{mass}
\Bigr[ -  p^{2}  +  m^{2} &-& \Pi_{\delta}(X) \Bigl] 
\Delta^{\gl }(X,p)  \nonumber \\
& = & {1 \over 2} \Big( \Pi^{\gl }(X,p) 
\big( \Delta^{+}(X,p)+ \Delta^{-}(X,p) \big) 
      + \big( \Pi^{+}(X,p) + \Pi^{-}(X,p) \big) 
\Delta^{\gl }(X,p) \Big) \nonumber \\
& + & {i \over 4} \Big\{ \Pi^>(X,p),\,\Delta^<(X,p) \Big\} 
 -  {i \over 4} \Big\{ \Pi^<(X,p),\, \Delta^>(X,p) \Big\} \;,
\end{eqnarray}
where we have used the identity (\ref{ra-><}) applied to the Green's
functions and self-energies. One recognizes Eq.~(\ref{trans}) as a transport 
equation while Eq.~(\ref{mass}) is called a mass-shell equation. We 
note that in the limit of free fields, Eqs.~(\ref{trans},\ref{mass}) become
\begin{eqnarray}\label{trans0}
p^{\mu} \partial_{\mu} \Delta^{\gl}_0(X,p) = 0 \;,
\end{eqnarray}
\begin{equation}\label{mass0}
\big[ p^{2} - m^{2} \big] \Delta^{\gl}_0(X,p) = 0 \;.
\end{equation}
Due to Eq.~(\ref{mass0}), $\Delta^{\gl}_0(X,p)$ is proportional to
$\delta (p^{2} - m^{2})$, and consequently free quasi-particles are always 
on mass-shell. We note that if the quasi-particle approximation 
(\ref{quasi-particle-approx}) were {\em not} used, the mass-shell equation 
would have the form
\be \label{off-shell-><}
\Bigr[ {1 \over 4} \partial^{2}  - p^{2}  
+  m^{2} \Bigl] \Delta^{\gl}_0(X,p) = 0 \;,
\ee
and the off-shell contribution to the Green's function $\Delta^{\gl}_0$ 
would be nonzero. 

The equations (\ref{trans},\ref{mass}) can be rewritten in a more compact 
way:
\begin{eqnarray}\label{trans1}
\Big\{ p^{2} - m^{2} + \Pi_{\delta}(X) + {\rm Re} \Pi^+(X,p),\, 
\Delta^{\gl }(X,p) \Big\}  &=&  i \Big( \Pi^>(X,p) \Delta^<(X,p) -
      \Pi^< (X,p) \Delta^> (X,p) \Big) \nonumber \\
& - & \Big\{ \Pi^{\gl}(X,p),\, {\rm Re} \Delta^+(X,p) \Big\}  \;,
\end{eqnarray}
\begin{eqnarray}\label{mass1}
\Bigr[ p^{2} - m^{2} + \Pi_{\delta}(X) + {\rm Re} \Pi^{+}(X,p) 
\Bigl] \Delta^{\gl }(X,p)  
&=& - \Pi^{\gl }(X,p) {\rm Re} \Delta^{+}(X,p) 
\nonumber \\
&-& {i \over 4} \Big\{ \Pi^>(X,p),\,\Delta^<(X,p) \Big\} 
 +  {i \over 4} \Big\{ \Pi^<(X,p),\, \Delta^>(X,p) \Big\} \;,
\end{eqnarray}
where
\be \label{Redelta}
{\rm Re} \Delta^{\pm}(X,p) \equiv {1 \over 2}
\Big(\Delta^+(X,p) + \Delta^-(X,p) \Big) \;,\;\;\;\;\;\;\;\;
{\rm Im} \Delta^{\pm}(X,p) \equiv  \pm {1 \over 2i}
\Big(\Delta^+(X,p) - \Delta^-(X,p) \Big) \;.
\ee  
We note that $\big(i\Delta^{\pm}(X,p) \big)^{\dagger} 
= -i \Delta^{\mp}(X,p)$ which is a direct consequence of the definitions 
(\ref{ret-xspace},\ref{adv-xspace}). The equation analogous to 
Eq.~(\ref{trans1}) was earlier derived in 
\cite{Kad62,Bez72,Mrowczynski:1997hy}. 

The gradient terms in the right-hand-sides of Eqs.~(\ref{trans1},\ref{mass1})
are usually neglected. The justification for dropping these terms is considered 
below. The small parameter that characterizes the gradient expansion will be 
denoted $\epsilon$. In Sec. \ref{collision-term} we will calculate the 
self-energies using a perturbative expansion in the coupling constant $g$. 
For the purposes of the discussion below, we will use a new  parameter to 
characterize the coupling constant expansion. This new parameter is defined 
by $\lambda = g$ for $\phi^{3}$ and $\lambda = \sqrt{g}$ for $\phi^{4}$ 
theory. The parameter $\lambda$ is introduced to simplify the notation: the 
leading order contributions to $\Pi_\delta(X)+{\rm Re}\Pi^+(X,p)$ and 
$\Pi^\gl(X,p)$ are of order $\lambda^{2}$ and $\lambda^{4}$ respectively, 
in both $\phi^{3}$ and $\phi^{4}$ theory \footnote{The different treatment 
of the $\phi^{3}$ and $\phi^{3}$ theory comes from the way the coupling 
constant enters the Lagrangian (\ref{lagran}). If we had defined interaction 
terms of the form $g^{2} \phi^{4}/4!$ and $g\phi^{3}/3!$ we would avoid using 
the parameter $\lambda$.}. 

Now, we compute the order of each term in Eq.~(\ref{trans1}). We can split 
the Poisson-like bracket on the left hand side into two pieces. 
The first piece does not contain the self-energy and is of order $\epsilon$. 
The second piece is of order $\lambda^{2} \epsilon$. The first term on the 
right hand side (the term without the Poisson-like bracket) is of order 
$\lambda^{4}$, and the term with the Poisson-bracket (the gradient term) 
is of order $\lambda^{4} \epsilon$. Thus, we find that the gradient term 
on the right hand side of Eq.~(\ref{trans1}) is of higher order than the 
remaining terms and can be dropped. Similarly, the gradient term on the 
right hand side of Eq.~(\ref{mass1}) is of order $\lambda^{4}\epsilon$ 
and it is of higher order than the remaining terms. 

We note that the argument presented above breaks down close to equilibrium. 
At equilibrium the non-gradient term on the right hand side of 
Eq.~(\ref{trans1}) is identically zero, and thus cannot be considered to 
be bigger than the gradient term. In the equilibrium case, it has been 
shown that the gradient term does not need to be dropped but can be combined 
with the interaction term on the left hand side to produce the usual Vlasov 
term \cite{Mrowczynski:1997hy}. The very-close-to-equilibrium situation is 
not fully understood.

\section{Equations for $\Delta^{\pm}$}
\label{eqs-+/-}

We also write down the transport and mass-shell equations satisfied 
by the retarded and advanced Green's functions. Starting with 
Eqs.~(\ref{DSpm1},\ref{DSpm2}), one finds
\begin{eqnarray}\label{trans-pm}
\Big\{ p^{2} - m^{2} + \Pi_{\delta}(X) + \Pi^{\pm}(X,p), 
\Delta^{\pm}(X,p) \Big\} = 0  \;,
\end{eqnarray}
\begin{eqnarray}\label{mass-pm}
\Bigr[ p^{2}  -  m^{2} + \Pi_{\delta}(X) + \Pi^{\pm}(X,p) \Bigl] 
\Delta^{\pm}(X,p)  =   1 \;.
\end{eqnarray}
We observe that the gradient terms drop out entirely in Eq.~(\ref{mass-pm}). 
The leading order corrections to this equation are second order in the 
gradient expansion.  Equation (\ref{mass-pm}) can be immediately solved 
to give
\begin{eqnarray}\label{Green-pm}
\Delta^{\pm}(X,p) = 
{1 \over  p^{2}  -  m^{2} + \Pi_{\delta}(X) + \Pi^{\pm}(X,p)}  \;.
\end{eqnarray}
We note that any function $f(K)$ satisfies the equation $\{ K, f(K) \} = 0$, 
for $K$ an arbitrary function of $X$ and $p$. Thus, the expression
(\ref{Green-pm}) solves Eq.~(\ref{trans-pm}) as well as Eq.~(\ref{mass-pm}). 

The real and imaginary parts of $\Delta^{\pm}$, which we will need later, are 
\begin{eqnarray}\label{Re-pm}
{\rm Re} \Delta^{\pm}(X,p) =
{p^{2}  -  m^{2} + \Pi_{\delta}(X) + {\rm Re}\Pi^+(X,p)
\over \big( p^{2}  -  m^{2} + \Pi_{\delta}(X) + {\rm Re}\Pi^+(X,p) \big)^{2}
+ \big({\rm Im}\Pi^+(X,p) \big)^{2}} \;,
\end{eqnarray}
\begin{eqnarray}\label{Im-pm}
{\rm Im} \Delta^{\pm}(X,p) =
{ \mp {\rm Im}\Pi^+(X,p)
\over \big( p^{2}  -  m^{2} + \Pi_{\delta}(X) + {\rm Re}\Pi^+(X,p) \big)^{2}
+ \big({\rm Im}\Pi^+(X,p) \big)^{2}} \;.
\end{eqnarray}  

Using the relations (\ref{ret-xspace},\ref{adv-xspace}), we can obtain 
the Wigner transforms of the retarded and advanced functions for free 
fields
\begin{equation}\label{pm0}
\Delta^{\pm}_0(X, p) = {1 \over p^{2} -m^{2} \pm ip_00^+} \;.
\end{equation}
Comparing the expressions (\ref{Green-pm}) and (\ref{pm0}), we find that 
in the limit $g \rightarrow 0$ the self-energies must satisfy
\begin{equation}\label{free-limit}
{\rm Im}\Pi^+(X,p) = - {\rm Im} \Pi^-(X,p) =
\left\{ \begin{array}{ccl} 
0^+ \;\;\; & {\rm for} &\;\; p_0 > 0 \;, \\
0^- \;\;\; & {\rm for} & \;\; p_0 < 0 \;. \\
\end{array} \right. 
\end{equation}

\section{Spectral Function}
\label{spectral}

In this section, we introduce another function which will be useful 
later on. The spectral function $A$ is defined as
\begin{equation}
\label{spec-Im}
A(x,y)  \buildrel \rm def \over = \langle [\phi (x), \phi (y)] \rangle 
= i\Delta^>(x,y)  -  i\Delta^<(x,y) 
= i\Delta^+(x,y)  -  i\Delta^-(x,y) 
= \mp 2 \, {\rm Im} \Delta^{\pm}(x,y) \;,
\end{equation}
where $[\phi (x), \phi (y)]$  denotes the field commutator.
For real fields, which are studied here, $A(X,p) = - A(X,-p)$.

From the transport and mass-shell equations (\ref{trans1}, \ref{mass1}), 
one immediately finds equations for $A(X,p)$ of the form
\begin{equation}\label{trans-spec}
\Big\{ p^{2} - m^{2} + \Pi_{\delta}(X) + {\rm Re} \Pi^+(X,p), \,
A(X,p) \Big\} = 2\Big\{ {\rm Im}\Pi^+(X,p),\,{\rm Re}\Delta^+(X,p) \Big\} \;,
\end{equation}
\begin{equation}\label{mass-spec}
\Bigr[ p^{2} - m^{2} + \Pi_{\delta}(X) + {\rm Re} \Pi^+(X,p) 
\Bigl] A(X,p)  
= 2 \, {\rm Im}\Pi^+(X,p) \, {\rm Re} \Delta^+(X,p) \;.
\end{equation}
Substituting the formula (\ref{Re-pm}) into the algebraic equation 
(\ref{mass-spec}), we find the solution 
\begin{eqnarray}\label{spec}
A (X,p) = { 2 {\rm Im}\Pi^+(X,p)
\over \big( p^{2}  -  m^{2} + \Pi_{\delta}(X) 
+ {\rm Re}\Pi^+(X,p) \big)^{2} + \big({\rm Im}\Pi^+(X,p) \big)^{2}} \;.
\end{eqnarray}
It is easy to show that the function (\ref{spec}) solves
Eq.~(\ref{trans-spec}) as well. In fact, the spectral function (\ref{spec})
can be found directly from Eq~(\ref{Im-pm}) due to the last equality in
Eq.~(\ref{spec-Im}).

The free spectral function can be obtained from the free retarded and
advanced functions (\ref{pm0}) by means of the relation (\ref{spec-Im}).
It can be also found from Eq.~(\ref{spec}) using the condition 
(\ref{free-limit}). In either case one needs to make use of the well-known 
identity 
\be \label{identity}
{1 \over x \pm i 0^+} = {\rm P}{1 \over x} \mp i \pi \delta (x) \;.
\ee
We obtain
\begin{eqnarray}\label{spec-free}
A_0 (X,p) = 2\pi \delta (p^{2} - m^{2} )
\bigr( \Theta (p_0) - \Theta (-p_0) \bigl) \;.
\end{eqnarray}

When ${\rm Im}\Pi^+ = 0$ the spectral function describes infinitely narrow 
quasi-particles with energies uniquely determined by their momenta. When 
${\rm Im}\Pi^+ \not= 0 $ the quasi-particles are of finite life time,
and the spectral function gives the energy distribution of a quasi-particle 
with a given momentum.

\section{Distribution Function}
\label{distri-f}

The distribution function $f(X,p)$ is defined through the equation
\begin{eqnarray}\label{def-f}
\Theta (p_0)\; A(X,p) \; f(X,p) \buildrel \rm def \over 
= \Theta (p_0) i \Delta^<(X,p)  \;,
\end{eqnarray}
where $A(X,p)$ is the spectral function (\ref{spec}). This definition is 
motivated by Eq.~(\ref{en-mom-free}) which leads us to identify $\Delta^\lg$ 
as the quantum analogue of the classical distribution function. We note that 
$f(X,p)$ depends not on the three-vector ${\bf p}$ but on the four-vector 
$p$. Using Eq.~(\ref{spec-Im}) and the identity (\ref{>vs<}) in the form 
$\Delta^<(X,p) = \Delta^>(X,-p)$, we have
\begin{eqnarray}\label{gr-f}
i \Delta^>(X,p)  =  \Theta (p_0) \; A(X,p) \; \big( f(X,p) + 1 \big)
- \Theta (-p_0) \; A(X,p) \; f(X,-p) \;,
\end{eqnarray}
\begin{eqnarray}\label{sm-f}
i \Delta^<(X,p)  = \Theta (p_0) \; A(X,p) \; f(X,p) 
- \Theta (-p_0) \; A(X,p) \; \big( f(X,-p) + 1 \big) \;.
\end{eqnarray}
In the case of free fields, Eqs.~(\ref{gr-f},\ref{sm-f}) simplify to
\begin{eqnarray}\label{sm-f0}
i \Delta^>_0(X,p) &=& 2\pi \: \delta (p^{2} - m^{2})
\Big(\Theta (p_0) \; \big( f_0(X,{\bf p}) + 1 \big)
+ \Theta (-p_0)\; f_0(X,-{\bf p}) \Big) \\ [1mm] \nonumber  
&=& {\pi \over E_p}\delta (E_p - p_0) 
\big( f_0(X,{\bf p}) + 1 \big)
+ {\pi \over E_p}\delta (E_p + p_0) f_0(X,-{\bf p}) \comma
\\ \label{gr-f0}
 i \Delta^<_0(X,p) &=& 2\pi \: \delta (p^{2} - m^{2})
\Big( \Theta (p_0) \; f_0(X,{\bf p})
+ \Theta (-p_0)\; \big( f_0(X,-{\bf p}) +1 \big) \Big) \\ [1mm] \nonumber
&=& {\pi \over E_p}\delta (E_p - p_0) f_0(X,{\bf p}) 
+ {\pi \over E_p}\delta (E_p + p_0) \big( f_0(X,-{\bf p}) + 1 \big) \;,
\end{eqnarray}
where the explicit form of the free spectral function (\ref{spec-free})
was used. 

For future use, we write down the results for the propagators $\Delta^c_0$, 
$\Delta^a_0$ and $\Delta^{\rm sym}_0$ in terms of the free distribution 
function $f_0$. Using the free retarded and advanced functions (\ref{pm0})
and the identity (\ref{identity}), we obtain
\ba 
\nonumber
i \Delta^c_0(X,p) &=&
  i \Delta^+_0(X,p) +  i \Delta^<_0(X,p) \\ \label{c-f0}
&=& {i \over p^{2} -m^{2} + i0^+} +  
2\pi \: \delta (p^{2} - m^{2})
\Big( \Theta (p_0) \; f_0(X,{\bf p})
+ \Theta (-p_0)\; f_0(X,-{\bf p}) \Big) 
\\[2mm] \nonumber
i \Delta^a_0(X,p) &=&
  -i \Delta^-_0(X,p) +  i \Delta^<_0(X,p) \\ \label{a-f0}
&=& {-i \over p^{2} -m^{2} - i0^+} +
2\pi \: \delta (p^{2} - m^{2})
\Big(\Theta (p_0) \; f_0(X,{\bf p}) 
+ \Theta (-p_0)\; f_0(X,-{\bf p}) \Big) \;,
\\ [2mm] \nonumber 
i \Delta^{\rm sym}_0(X,p) &=& 
  i \Delta^>_0(X,p) +  i \Delta^<_0(X,p) \\ [2mm] \label{sym-f0}
&=& 2\pi \: \delta (p^{2} - m^{2})
\Big( 2\Theta (p_0) \;f_0(X,{\bf p})
+ 2\Theta (-p_0)\; f_0(X,-{\bf p}) + 1 \Big) \;.
\ea

We note that in the case of equilibrium (homogeneous) systems, the familiar 
definition of the distribution function is 
\be \label{dis-eq}
f^{\rm eq}({\bf p}) \buildrel \rm def \over = 
{1 \over V} \langle a^{\dagger}({\bf p}) a({\bf p}) \rangle \;,
\ee
where $V$ denotes the system volume and $a({\bf p})$ and 
$a^{\dagger}({\bf p})$ are annihilation and creation operators,
respectively. The definition (\ref{def-f}) should reduce to (\ref{dis-eq}) 
when an equilibrium (homogeneous) system is considered. We discuss this point 
below for the case of a noninteracting system for which the field that solves 
the equation of motion (\ref{motion}) can be written as
\be 
\label{field-sol}
\phi(x) =   \int {d^3k \over \sqrt{(2\pi )^{3} 2 E_k }}
\Big( e^{-ikx} \; a({\bf k}) + e^{ikx} \; a^{\dagger}({\bf k}) \Big) \;, 
\ee
where $k \equiv (E_k, {\bf k})$. Substituting the solution (\ref{field-sol}) 
into the definition (\ref{contour-parts}) one finds
\ba \label{>-eq}
i\Delta^>_0(X,p) &=& \int  {d^3k \over \sqrt{(2\pi )^{3} 2 E_k }} \;
 {d^3q \over \sqrt{(2\pi )^{3} 2 E_q }} \; (2\pi)^{4} \\[2mm] \nonumber
&\times&\Big[  e^{-i(k+q)X} \delta^{(4)}\Big(p - {k-q \over 2} \Big)
                 \langle a({\bf k}) a({\bf q}) \rangle 
+  e^{-i(k-q)X} \delta^{(4)}\Big(p - {k+q \over 2} \Big) 
                 \langle a({\bf k}) a^{\dagger}({\bf q}) \rangle 
\\[2mm]\nonumber 
&&+  e^{i(k-q)X} \delta^{(4)}\Big(p + {k+q \over 2} \Big)
                 \langle a^{\dagger}({\bf k}) a({\bf q}) \rangle 
+ \; e^{i(k+q)X} \delta^{(4)}\Big(p + {k-q \over 2} \Big)
         \langle a^{\dagger}({\bf k}) a^{\dagger}({\bf q}) \rangle \Big] \;.
\ea
Since we restrict ourselves to the consideration of homogeneous 
systems, we require that the function $\Delta^>_0(X,p)$ is independent of 
$X$. One ansatz that satisfies this constraint is given by 
\be 
\label{condition1}
\langle a({\bf k}) a({\bf q}) \rangle = 0 \;,\;\;\;\;\;\;\;\;\;\;\;
\langle a^{\dagger}({\bf k}) a^{\dagger}({\bf q}) \rangle = 0 \;, 
\ee
and
\be \label{condition2} 
 \langle a({\bf k}) a^{\dagger}({\bf q}) \rangle 
= {\delta^{(3)} ({\bf k} -{\bf q}) \over V}
 \langle a({\bf k}) a^{\dagger}({\bf k}) \rangle \;, \;\;\;\;\;\;\;\;\;\;\;
 \langle a^{\dagger}({\bf k}) a({\bf q}) \rangle
= {\delta^{(3)} ({\bf k} -{\bf q}) \over V}
 \langle a^{\dagger}({\bf k}) a({\bf k}) \rangle \;.  
\ee
Using these expressions, the integrals in Eq.~(\ref{>-eq}) are trivially 
performed and we obtain 
\be \label{>-eq-2}
i\Delta^>_0(p) = {\pi \over E_p}\delta (E_p - p_0)\;
 {1 \over V} \langle a({\bf p}) a^{\dagger}({\bf p}) \rangle \; 
+ {\pi \over E_p}\delta (E_p + p_0) \;  {1 \over V}
 \langle a^{\dagger}(-{\bf p}) a(-{\bf p}) \rangle \;.
\ee
Using the commutation relation 
$[a({\bf p}), a^{\dagger}({\bf k})] =  \delta^{(3)} ({\bf p} -{\bf k})$,
which gives
$$
 \langle a({\bf k}) a^{\dagger}({\bf k}) \rangle 
=  \langle a^{\dagger}({\bf k}) a({\bf k}) \rangle + V \;,
$$
one finds that Eq.~(\ref{>-eq-2}) combined with the definition (\ref{dis-eq}) 
reproduces (\ref{gr-f0}). Thus we have shown that the definitions 
(\ref{def-f}) and (\ref{dis-eq}) are consistent with each other for 
homogeneous systems. One notes that for nonhomogeneous systems the 
conditions (\ref{condition1},\ref{condition2}) are, in principle, not 
fulfilled. As a result, the functions $\Delta^{\lg}_0(X,p)$ are not only 
$X-$dependent but also have support for off-shell momenta $p$ in
agreement with Eq.~(\ref{off-shell-><}). Therefore, the expressions
(\ref{sm-f0},\ref{gr-f0}) can be treated as a representation of 
$\Delta^{\lg}_0(X,p)$ only for weakly nonhomogeneous systems.

\section{Mass-shell constraint}
\label{mass-shell-con}

In this section we discuss two issues that are related to the mass shell 
constraint. We first show that in the homogeneous limit (or neglecting
gradient terms), the mass shell condition is satisfied if the distribution 
function is defined in terms of the spectral function as in Eq.~(\ref{def-f}), 
and if this distribution function satisfies the transport equation. 
Secondly, we show that using this definition of the distribution function, 
the mass-shell constraint reduces to the familiar definition of the mass-shell 
condition on the four-momentum in the limit of zero width quasi-particles. 

The transport and mass-shell equations for a homogeneous system, which are
obtained from Eqs.~(\ref{trans1}) and (\ref{mass1}) by dropping gradient
terms, read
\begin{eqnarray}
\label{zero-grad}
\Pi^>(X,p) \Delta^<(X,p) - \Pi^< (X,p) \Delta^> (X,p) &=& 0 \;,\\[2mm]
\Bigr[ p^{2} - m^{2} + \Pi_{\delta}(X) + {\rm Re} \Pi^+(X,p) \Bigl]
\Delta^{\gl }(X,p)  &=& - \Pi^{\gl}(X,p) {\rm Re} \Delta^+(X,p) \;.
\end{eqnarray}
These equations can be rewritten using Eqs.~(\ref{gr-f},\ref{sm-f})
which express $\Delta^<(X,p)$ through the distribution function.
We work below with the mass-shell equation for $\Delta^<(X,p)$ and 
restrict to $p_0 > 0$. It is straightforward to repeat all steps with 
$p_0 < 0$, and for $\Delta^<$ with $p_0 > 0$ and $p_0 < 0$. We 
find
\ba
\label{tran-0-f}
A(x,p)
\bigg[ \Pi^>(X,p) \, f(X,p) - \Pi^< (X,p) \, \Big( f(X,p)+1\Big) \bigg]
&=& 0\,, \\
\label{mass-0-f-1}
\Bigr[ p^{2} - m^{2} + \Pi_{\delta}(X) + {\rm Re} \Pi^+(X,p)
\Bigl] \, A(X,p) \, f(X,p)
&=& - i \Pi^<(X,p) {\rm Re} \Delta^+(X,p) \,.
\ea
Using the spectral function equation (\ref{mass-spec}), the mass-shell
equation (\ref{mass-0-f-1}) has the form
\begin{equation}\label{mass-0-f-2}
{\rm Re}\Delta^+(x,p) \;
\bigg[ \Pi^>(X,p) \, f(X,p) 
- \Pi^< (X,p) \, \Big( f(X,p)+1\Big) \bigg] = 0 \;.
\end{equation}
One sees that if $f$ solves Eq.~(\ref{tran-0-f}), it automatically 
satisfies Eq.~(\ref{mass-0-f-2}). Equivalently, the transport and 
mass-shell equations (\ref{trans1}) and (\ref{mass1}) can be replaced 
with the two equations (\ref{mass-spec}) and (\ref{tran-0-f}) in the 
homogeneous limit.

We note that the equations (\ref{tran-0-f}, \ref{mass-0-f-1}), 
which were obtained by neglecting gradient terms, are exact 
for equilibrium (homogeneous) systems. These equations can be 
written as
\ba
\label{kms-pi}
\Pi^>(p) \, f^{\rm eq}(p) =
 \Pi^<(p) \Big(f^{\rm eq}(p) + 1 \Big) \;,
\ea
which is the well-known Kubo-Martin-Schwinger (KMS) condition, 
see {\it e.g.} \cite{Kad62}, that is satisfied by the self-energy in 
equilibrium. We also note that in going from equations (\ref{trans1}) 
and (\ref{mass1}) to equations (\ref{mass-spec}) and (\ref{tran-0-f}), 
we have replaced four equations with two. This reduction in the number 
of independent Green's functions is expected since in equilibrium 
the two functions $\Delta^\lg(X,p)$ are not independent but related 
through a KMS condition of the form (\ref{kms-pi}).

We have shown above that when the distribution function is defined by 
Eq.~(\ref{def-f}), the transport and mass-shell equations 
(\ref{trans1},\ref{mass1}) can be replaced by the equations (\ref{mass-spec}) 
and (\ref{tran-0-f}). In the limit ${\rm Im}\Pi^+ \ra 0$ the solution to 
Eq.~(\ref{mass-spec}), which is given by Eq.~(\ref{spec}), becomes 
\be \label{spec2}
A(X,p) = 2\pi(\Theta(p_0)-\Theta(-p_0)) \;
\delta( p^{2}  -  m^{2} + \Pi_{\delta}(X) + {\rm Re}\Pi^+(X,p) ) \;.
\ee
The argument of the delta function gives the usual form of the mass-shell 
constraint of the four-momentum
\be
\label{dis-rel2}
p^2  -  m^2 + \Pi_{\delta}(X) + {\rm Re}\Pi^+(X,p) = 0 \;,
\ee
which tell us that only three out of four momentum components are 
independent. For ${\rm Im}\Pi^+$ small but finite, this equation determines 
the position of the maximum of the spectral function. 

We note that when the system of interest is significantly
inhomogeneous, {\it cf.} Eq.~(\ref{off-shell-><}), or when an interaction
generates a non-negligible value of ${\rm Im}\Pi^+$, the function
$\Delta^{\gl}(X,p)$, which solves Eq.~(\ref{mass1}), has support
from momenta not satisfying the relation (\ref{dis-rel2}). In this
case, one has finite-width quasi-particles and the equation (\ref{dis-rel2})
only gives, according to Eq.~(\ref{spec}), the most probable energy of
a quasi-particle. Thus, the statement that the four-momentum is on
the mass-shell (\ref{dis-rel2}) and the statement that the function 
$\Delta^{\gl}(X,p)$ satisfies the mass-shell constraint (\ref{mass1}) 
are, in general, not equivalent.

\section{Transport equation}
\label{trans-eq}

The distribution function $f$ satisfies the transport equation
which can be obtained from Eq.~(\ref{trans1}) for $\Delta^>$ or $\Delta^<$. 
Using Eqs.~(\ref{trans-spec}) and (\ref{def-f}), one finds
\begin{eqnarray}
\label{trans-f-1}
\Theta(p_0) A(X,p)\,\Big\{ p^{2} - m^{2} 
+ \Pi_{\delta}(X) &+& {\rm Re} \Pi^+(X,p) , \, f(X,p) \Big\} 
\nonumber \\
&=& i \Theta(p_0)\Big[ A(X,p) \,\Big( \Pi^>(X,p) \, f(X,p) -
      \Pi^< (X,p) \, \big(f(X,p)+1) \Big) \nonumber \\
&+& f(X,p) \, \Big\{ \Pi^>(X,p),\, {\rm Re} \Delta^+(X,p) \Big\} 
\nonumber \\
&-&  \big(f(X,p)+1\big)\,\Big\{ \Pi^<(X,p),\,{\rm Re} \Delta^+(X,p) 
\Big\}\Big]\;,
\end{eqnarray}
where we have used the identity
$$
\big\{A ,\, B\,C \big\} = \big\{A ,\, B \big\}\,C 
+ \big\{A ,\, C \big\}\,B \;.
$$
Neglecting the gradient terms in the r.h.s. of Eq.~(\ref{trans-f-1}),
we obtain the equation
\be \label{trans-f-2}
\Theta(p_0) \Big\{ p^{2} - m^{2} + \Pi_{\delta}(X) + {\rm Re} \Pi^+(X,p) , 
\, f(X,p) \Big\} 
= i \Theta(p_0)\Big( \Pi^>(X,p) \, f(X,p) 
-  \Pi^< (X,p) \, \big(f(X,p)+1)\Big) \;.
\ee

The transport equation (\ref{trans-f-2}) greatly simplifies for
zero-width quasi-particles.  Using the mass-shell condition (\ref{dis-rel2}),
and taking the positive energy solution, one can evaluate the Poisson-like 
bracket to obtain 
\be \label{trans-f-3}
\Theta(p_0) \,
\Big\{ p^{2} - m^{2} + \Pi_{\delta}(X) + 
{\rm Re} \Pi^+(X,p) , \, f(X,{\bf p}) \Big\} 
= 2 E_p \Big({\partial \over \partial t} + 
{\bf v} \cdot \nabla \Big) f(X,{\bf p})
+ 2 \nabla V(X, p) \, \nabla_p f(X,{\bf p}) \;,
\ee
where  
$$
V(X,p) \equiv \Pi_{\delta}(X) + {\rm Re} \Pi^+(X,p) \;,
$$
and the velocity ${\bf v} \equiv \partial E_p /\partial {\bf p}$. Note 
that $E_p$ is not $\sqrt{{\bf p}^2 + m^2}$ but the positive solution of 
Eq.~(\ref{dis-rel2}). Substituting Eq.~(\ref{trans-f-3}) into 
Eq.~(\ref{trans-f-2}), the transport equation takes the form
\ba \label{trans-f}
E_p  \Big({\partial \over \partial t} + 
{\bf v} \cdot \nabla \Big) f(X,{\bf p})
+ \nabla V(X,{\bf p}) \, \nabla_p f(X,{\bf p})
=\Theta(p_0) \, \frac{i}{2} \, \big[ \Pi^>(X,p) \, f(X,{\bf p}) 
-  \Pi^< (X,p) \, \big(f(X,{\bf p})+1)\big] \;.
\ea
In the left hand side of Eq.~(\ref{trans-f}), one recognizes the 
standard drift and Vlasov terms \cite{Gro80}. The calculation of 
the self-energies ${\rm Re}\Pi^+$ and $\Pi_\delta$ in the Vlasov 
term is straightforward and the physical interpretation is well known: 
they contribute to mean field effects and result in the generation of 
an effective mass. In this paper we will not discuss these terms 
(they have been studied in detail in \cite{Mrowczynski:1989bu}). 
From the right hand side of Eq.~(\ref{trans-f}), one defines the 
collision term $C(X,p)$ as
\ba
\label{coll-1}
\Theta(p_0) C(X,p) \buildrel \rm def \over = 
\Theta(p_0) \, \frac{i}{2} \, \big[ \Pi^>(X,p) \, f(X,{\bf p}) 
-  \Pi^< (X,p) \, \big(f(X,{\bf p})+1)\big] \;.
\ea
The structure of this collision term is the subject of the next five 
sections.

\section{Collision Term Structure}
\label{collision-term}

In order to obtain physical results from the transport equation derived 
in the previous sections, we must specify the self-energies that appear 
in the collision term. We will assume that the interaction is sufficiently 
weak that the self-energies can be expanded in powers of the coupling 
constant, and thus expressed through the Green's functions. We will develop 
a diagrammatic expansion of the self-energies and identify the different 
physical processes that contribute to the collision term (\ref{coll-1}). 
The collision terms we obtain are local in space-time, and consequently 
valid in the Markovian limit where all memory effects are neglected.
 
We first rewrite Eq.~(\ref{coll-1}) by multiplying by the spectral 
function $A(X,p)$ and using Eq.~(\ref{def-f}). Thus, we obtain the equation
\begin{eqnarray}
\label{coll-term1}
\Theta(p_0) A(X,p) \: C(X,p) 
 = \Theta(p_0)\frac{1}{2}\Big( \Pi^< (X,p) \Delta^> (X,p) 
-\Pi^>(X,p) \Delta^<(X,p) \Big) \;.
\end{eqnarray}

Evolution along the contour illustrated in Fig.~\ref{fig-contour} 
is formally very similar to evolution along the real time axis, and 
consequently it is possible to define perturbation theory on the contour. 
Self-energies with real time arguments, in particular $\Pi^\gl$, have to 
be extracted from the contour self-energy. Since these calculations involve 
summations over contributions from both branches of the contour, they are 
much more difficult than their vacuum counterparts. There are several 
different methods that can be used which are based on different 
representations of real time field theory \cite{Gelis:1997zv} and each 
has its advantages and proponents. At the lower loop levels, the relevant 
self-energies have been calculated in several different ways both in 
equilibrium \cite{Weldon:jn} and non-equilibrium \cite{Mrowczynski:1989bu}.
(A general structure of the equilibrium self-energy expressed through
the scattering amplitude has been studied in \cite{Jeon:1998zj}.) At three 
or more loops, the calculation is extremely cumbersome and requires the 
use of special methods. 

Our calculation of the self-energies, which enter the collision 
term of transport equation, uses the Keldysh representation formulated 
in terms of retarded, advanced and symmetric Green functions. To 
understand why the Keldysh representation is better suited to 
perform such a calculation, we need to look at the Wigner transformed 
propagators for non-interacting fields in the Keldysh and 1/2 
representation. In the 1/2 basis the propagators are given by equations 
(\ref{sm-f0}, \ref{gr-f0}, \ref{c-f0}, \ref{a-f0}) and in the 
Keldysh basis by (\ref{pm0}, \ref{sym-f0}). The functions 
$i\Delta_0^<$ and $i\Delta_0^>$ do not contain off-mass-shell
contributions that have the form of propagators. They are non-zero 
only on the mass-shell. In contrast, $i\Delta_0^c$ and $i\Delta_0^a$ 
contain one piece that corresponds to time-ordered and anti-time-ordered 
propagation, respectively, and one piece that is non-zero only on 
the mass shell. In the Keldysh basis, however, none of the Green's
functions mixes the propagating pieces and terms that are non-zero 
only on the mass-shell: $\Delta_0^+$ and $\Delta_0^-$ contain only
the contributions corresponding to retarded and advanced propagation, 
respectively, and $\Delta_0^{\rm sym}$ is non-zero only on-mass-shell.
Remembering that on-mass-shell Green functions represent real
particles and propagating Green functions virtual particles, it
is not unexpected that this clean separation of real and virtual 
particles makes the physical interpretation of the collision 
term much more straightforward. 

We switch to the Keldysh basis using the relations (\ref{ra-><}) 
and  analogous relations for the self-energies. The collision term 
(\ref{coll-term1}), which will be computed in the next sections 
up to four-loop level, then equals
\ba
\label{KER}
\Theta(p_0) A(p) \: C(p) =\frac{1}{4} \Theta(p_0) 
\Big[ \Pi^{\rm sym}(p) \big(\Delta^+(p) -\Delta^-(p)\big)
-  \big(\Pi^+(p)-\Pi^-(p) \big) \Delta^{\rm sym}(p) \Big] \;.
\ea

It will be shown that in the Keldysh representation, all contributions 
to the collision term have the form of `cuts' \cite{Carrington:2002bv}. 
We use the word `cut' in the way that it is usually used when discussing 
the calculation of the imaginary part of a diagram: a `cut line' passes 
through a diagram from top to bottom dividing it into two parts, so that 
each part contains at least one external leg.  A propagator that is crossed 
by the cut line becomes a `cut propagator' which means that it has been put 
on the mass shell.  

We introduce the notion `central cut' to refer to the cut for which 
all loops contain at least one cut propagator. For the one-loop 
diagram in $\phi^3$ theory, and the two-loop diagram in $\phi^4$ theory, 
the only possible cut is the central cut. At two and more loops in $\phi^3$ 
theory, and three and more loops in $\phi^4$ theory, all diagrams contain 
contributions from non-central cuts. In addition, there are some diagrams 
for which no central cut exists. In each case however, the contributions 
from non-central cuts can be included as renormalized lower loop terms. 
An example of this is presented in Sec.~\ref{phi3-2L}. The main thrust 
of our paper involves the analysis of the central cuts of three- and 
four-loop diagrams. 

The calculation of a self-energy graph involves two non-trivial steps: 
the calculation of the integrand and the computation of the integral 
itself. In this paper we are primarily interested in the calculation of 
integrands which determine the structure of the collision term. At three 
or more loops, calculating by hand in any representation is extremely 
tedious. We use a {\it MATHEMATICA} program developed by one of us, which 
is described in detail in \cite{Carrington:jt}, to calculate the integrand. 
To use the program one assigns momenta and indices to every line and 
vertex in the diagram. Each index can take the values 1 or 2, corresponding 
to the top and bottom branches of the contour. The indices that correspond 
to external legs take fixed values, and a specific combination corresponds 
to a specific self-energy in the 1/2 representation, in analogy with 
Eq.~(\ref{1/2}). Internal indices are summed over their two possible values 
1 and 2. Using Eqs.~(\ref{1/2},\ref{sym-xspace},\ref{ra-><}), one switches 
to the Keldysh representation. A huge number of terms is produced even 
for fairly simple diagrams. However, there are many cancellations between 
these terms. The program identifies the surviving contributions and provides 
them as output. 

In Secs.~\ref{one-loop} and \ref{two-loop} we rederive the well 
known one- and two-loop results to illustrate our technique. In 
Sec.~\ref{three-loop} we use our method to obtain results for 
three-loop diagrams which contribute to $2\leftrightarrow 3$ processes 
in $\phi^3$ theory, and in Sec.~{\ref{four-loop} we study four-loop 
diagrams in $\phi^4$ theory which correspond to $2\leftrightarrow 4$ 
and $3\leftrightarrow 3$ processes.

\section{One-loop contributions}
\label{one-loop}

We start by considering the lowest order contributions. In the $\phi^4$ model,
the only one-loop diagram is the tadpole. It is well known that this diagram 
only contributes to mean-field dynamics, see {\it e.g.} \cite{Mrowczynski:1989bu}, and therefore it will not be discussed here. 
For $\phi^3$ theory, the one-loop contribution to the self-energy, which 
is shown in Fig.~\ref{fig-one-loop}, has a non-trivial imaginary part. 
The related scattering amplitude corresponds to 1 $\leftrightarrow$ 2 
processes which are kinematically forbidden for on-mass-shell particles, 
as an on-shell particle of mass $m$ cannot decay into two on-shell particles 
with the same masses. However, it is important to remember that the process 
is allowed for virtual particles and has been repeatedly discussed, in 
particular, in the context of electron-positron-photon interactions. We will 
discuss this contribution in detail to illustrate the method that we will 
use in the analysis of more complicated multi-loop diagrams. 
 
At this point, we introduce some notational simplifications. We suppress 
the $X-$dependence of all functions. All Green's functions used in our 
perturbative calculations correspond to free fields, and thus we suppress 
the index `0' on these functions. Similarly, we suppress from now on the 
index `0' which has been used to denote the free distribution function 
and the free spectral function. Finally, we define
on-shell four-momenta
\ba
\label{on-shell}
\tilde p \equiv (E_p,{\bf p})\;,
\ea
with $E_p \equiv \sqrt{{\bf p}^2 +m^2}$. We note that $\sqrt{{\bf p}^2 +m^2}$
should be denoted $E_p^0$, but as above the index `0' is suppressed.

The Keldysh self-energies corresponding to the diagram from 
Fig.~\ref{fig-one-loop} are
\ba
\label{parts}
&& \Pi^+(p) = i\frac{g^2}{2}\int \frac{d^4q}{(2\pi)^4}[\Delta^-(q)
\Delta^{\rm sym}(q+p) + \Delta^{\rm sym}(q)\Delta^+(q+p)] \comma
&& \Pi^-(p) = i\frac{g^2}{2}\int \frac{d^4q}{(2\pi)^4}[\Delta^{\rm sym}(q)
\Delta^-(q+p) + \Delta^+(q)\Delta^{\rm sym}(q+p)] \comma
&& \Pi^{\rm sym}(p) =  i\frac{g^2}{2} \int \frac{d^4q}{(2\pi)^4}[
\Delta^{\rm sym}(q)\Delta^{\rm sym}(q+p) + \Delta^+(q)\Delta^-(q+p)
+ \Delta^-(p)\Delta^+(q+p) ]\per
\ea
We substitute these self-energies into the right hand side 
of the collision term (\ref{KER}) and express $\Delta^{\rm sym}(p)$ 
through the distribution function according to Eq.~(\ref{sym-f0}). 
We also use Eqs.~(\ref{spec-Im}, \ref{spec-free}) which give
\ba
\label{pm0-2}
i\big(\Delta^+(p)-\Delta^-(p)\big) = 
{\rm sgn}(p_0)\, 2\pi \,\delta(p^2-m^2)\;,
\ea
and we observe that 
\ba
\label{pole}
\int \frac{d^4q}{(2\pi)^4}\Delta^+(q)\Delta^{+}(q+p) 
= \int \frac{d^4q}{(2\pi)^4}\Delta^-(q)\Delta^{-}(q+p) = 0 \;,
\ea
due to the positions of poles of $\Delta^\pm$.
Thus, we obtain
\ba
\Theta (p_0) \: A(p) \: C_{1L-\phi^3}(p) =  
\Theta (p_0) \: \frac{g^2}{16} \int  \frac{d^4q}{(2\pi)^4} 
&& \sum_{\{n_q,n_p,n_r\}=\pm1} 
{\cal F}_3(n_p,{\bf p}\, ; n_q,{\bf q}\, ;-n_r,-{\bf r}) \\ [2mm]
&& \times
\prod_{\alpha = \{p,\, q,\,r\}} 
n_{\alpha}\,{\rm sgn}(\alpha_0)\: 2\pi
\delta(\alpha^2-m^2)\,\Theta(n_{\alpha}\alpha_0) \;, \nonumber
\ea 
where $r \equiv q +p$ and  $\alpha$ is a generic four-momentum variable
which equals $p$, $q$ or $r$, and $n_\alpha = \pm 1$. The statistical 
factor ${\cal F}_N$ is defined as
\ba
\label{curleyF}
{\cal F}_N(n_1,{\bf p}_1;n_2,{\bf p}_2; \cdots n_N, {\bf p}_N) 
&\equiv&  
\Omega(n_1,{\bf p}_1) \, \Omega(n_2, {\bf p}_2) \, \cdots
\Omega(n_N,{\bf p}_N) 
\\ [2mm] \nonumber
&-& \Omega(-n_1,-{\bf p}_1)\, \Omega(-n_2,-{\bf p}_2) \, 
\cdots \Omega(-n_N, -{\bf p}_N) 
\ea
with  
\ba
\label{omega}
\Omega(n,{\bf p}) \equiv (1+n)\big(1 + f({\bf p})\big)
+(1-n) \, f(-{\bf p}) \;.
\ea
We note that the asymmetry in the signs of $(n_q, {\bf q})$ 
and $(n_r, {\bf r})$ in the factor 
${\cal F}_3(n_p,{\bf p}\, ; n_q,{\bf q}\, ;-n_r,-{\bf r})$ 
occurs because the momenta $q$ and $r$ are defined to flow 
in opposite directions in Fig.~\ref{fig-one-loop}.

\par\begin{figure}[H]
\begin{center}
\includegraphics[width=5cm]{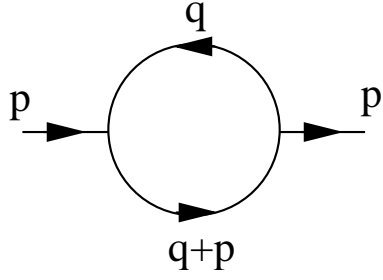}
\end{center}
\caption{\label{fig-one-loop}
One-loop contribution to the $\phi^3$ self-energy.}
\end{figure}

To further simplify our result, we write the delta functions as
\be
{\rm sgn}(\alpha_0)\: \delta(\alpha^2-m^2)
= {1 \over 2E_\alpha} \sum_{n_\alpha = \pm 1} n_\alpha \,
\Theta(n_\alpha \alpha_0)\,
\delta(\alpha_0 - n_\alpha E_\alpha) \;,
\ee
where $E_\alpha \equiv \sqrt{m^2 + \bold{\alpha}^2}$,
and get   
\ba \label{one-loop-coll-term}
\Theta (p_0) \: A(p) \: C_{1L-\phi^3}(p) = 
\Theta (p_0) \: \frac{g^2 \pi^3}{16} \int \frac{d^4q}{(2\pi)^4}
\sum_{\{n_p,n_q,n_r\}=\pm1} 
{\cal F}_3(n_p,{\bf p}\, ; n_q,{\bf q}\, ;-n_r,-{\bf r})
\prod_{\alpha = \{p,\, q,\,r\}} 
{1 \over E_\alpha} \: \delta(\alpha_0 - n_\alpha E_\alpha) \,.
\ea
We rewrite the expression (\ref{one-loop-coll-term}) by introducing 
the delta  function $\delta^{(4)}(p+q-r)$ and an integral over $r$, 
which allows us to treat $r$ as 
an independent variable. Performing the integrals over $p_0$ and 
$r_0$, one obtains
\ba
\label{1TO2}
\Theta (p_0) \: A(p) \: C_{1L-\phi^3}(p) &=& 
\Theta (p_0) \: \frac{g^2 \pi}{64} 
\int \frac{d^3q}{(2\pi)^3} \int \frac{d^3r}{(2\pi)^3}
\: (2\pi)^3\delta^{(3)}({\bf p}+{\bf q}-{\bf r})
\prod_{\alpha = \{p,\, q,\,r\}} 
{1 \over E_\alpha}
\\[2mm] \nonumber
&\times& \sum_{\{n_p,n_q,n_r\}=\pm1} 
\delta(p_0 - n_p E_p) \;
2\pi \delta(n_p E_p+n_q E_q-n_r E_r) \;
{\cal F}_3(n_p,{\bf p}\, ; n_q,{\bf q}\, ;-n_r,-{\bf r})
\ea

We discuss below the physical interpretation of the positive energy
piece of the collision term (\ref{1TO2}) which will be denoted with
the index $(+)$. We drop the theta function in Eq.~(\ref{1TO2}) 
and take only the $n_p=1$ term in the sum. We also use the explicit form 
of the free spectral function (\ref{spec-free}) to make the replacement
\be
\label{pos-only}
\Theta(p_0)\, A(p) = \frac{\pi}{E_p} \:\delta(p_0 - E_p) \;.
\ee
In principle, we are left with four terms that come from the sum over 
$n_q$ and $n_r$ in Eq.~(\ref{1TO2}). One of these terms corresponds to 
a delta function of the form $\delta(E_p+E_q+E_r)$ which has no support. 
The remaining three terms correspond to the processes symbolically denoted 
as: $p \leftrightarrow q + r$, $p + q \leftrightarrow r$ and 
$p + r \leftrightarrow q $. The last two terms can be combined by making 
the variable transformation $q \leftrightarrow r$ in the second term. 

We consider first the contribution to the collision term (\ref{1TO2}) 
obtained from choosing $n_q=-1$ and $n_r=1$. Making the change of variables 
${\bf q}\ra -{\bf q}$, we find
\ba
C^{(+)}_{1L-\phi^3}[p\leftrightarrow q+r] = 
\frac{g^2}{64} \int \frac{d^3q}{(2\pi)^3E_q}
\int \frac{d^3r}{(2\pi)^3E_r}
\;(2\pi)^4 \delta^{(4)}(\tilde p - \tilde q - \tilde r)
\; {\cal F}_3(1,{\bf p}\, ; -1,-{\bf q}\, ;-1,-{\bf r}) \;, 
\ea
where the statistical factor equals
\ba
\label{12}
{\cal F}_3(1,{\bf p}\, ; -1,-{\bf q}\, ;-1,-{\bf r})=
2^3 \Big[ \big(1+f({\bf p})\big) f({\bf q})
f({\bf r}) - f({\bf p}) \big( 1 +f({\bf q}) \big)
\big(1+f({\bf r}) \big) \Big] \;.
\ea
This process is represented in Fig. \ref{1to2-B}.

\par\begin{figure}[H]
\begin{center}
\includegraphics[width=3cm]{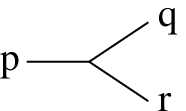}
\end{center}
\caption{$p \leftrightarrow q + r$  scattering process.}
\label{1to2-B}
\end{figure}

\par\begin{figure}[H]
\begin{center}
\includegraphics[width=3cm]{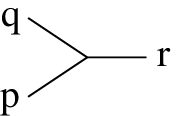}
\end{center}
\caption{$p+q \leftrightarrow r$ scattering process.}
\label{2to1-B}
\end{figure}

Next, we look at the contributions to  Eq.~(\ref{1TO2}) obtained from 
choosing  $n_q=n_r=1$ and $n_q=n_r=-1$. After changing variables 
and combining the two terms, we get
\ba
C^{(+)}_{1L-\phi^3}[p+q \leftrightarrow r] = 
\frac{g^2}{32} 
\int \frac{d^3q}{(2\pi)^3E_q}
\int \frac{d^3r}{(2\pi)^3E_r}
\;(2\pi)^4 \delta^{(4)}(\tilde p + \tilde q - \tilde r)
{\cal F}_3(1,{\bf }q\, ; 1,{\bf q}\, ;-1,-{\bf r}) \;,
\ea
with
\ba
\label{21}
{\cal F}_3((1,{\bf p}\, ; 1,{\bf q}\, ;-1,-{\bf r})=
2^3\Big[ \big(1+f({\bf p})\big) \big(1+f({\bf q})\big)
f({\bf r}) - f({\bf p}) f({\bf q})
\big(1+f({\bf r}) \big) \Big] \;.
\ea
This contribution is illustrated in Fig.~\ref{2to1-B}.

The complete one-loop collision term equals
\ba \label{1-loop-CT}
C^{(+)}_{1L-\phi^3}(p) &=& 
\int \frac{d^3q}{(2\pi)^3E_q}
\int \frac{d^3r}{(2\pi)^3E_r} \: |{\cal M}|^2_{1L-\phi^3}
\\[2mm] \nonumber
&\times&\bigg[
(2\pi)^4 \delta^{(4)}(\tilde p - \tilde q - \tilde r)
\; \frac{1}{2!} \;
\Big[ \big(1+f({\bf p})\big) f({\bf q}) f({\bf r})
- f({\bf p}) \big( 1 +f({\bf q})\big) \big(1+f({\bf r}) \big) \Big] 
\\ [2mm] \nonumber 
&+& ~(2\pi)^4 \delta^{(4)}(\tilde p + \tilde q - \tilde r) \; 
\Big[\big(1+f({\bf p})\big) \big(1+f({\bf q})\big) f({\bf r})
- f({\bf p}) \: f({\bf q}) \big(1+f({\bf r}) \big) 
\Big] \Bigg] \;,
\ea
where the scattering matrix element equals 
$|{\cal M}|^2_{1L-\phi^3}=g^2/4$. The two terms proportional to 
$f({\bf p})$ correspond to `loss' contributions and the terms 
proportional to $(1 + f({\bf p}))$ correspond to `gain' contributions. 
In the first term, which represents the process $p\leftrightarrow q+r$,
there is the factor $1/2!$ because of the integration over momenta of 
two identical particles in either final or initial scattering state. 

We remind the reader that the contribution (\ref{1-loop-CT}) to the 
collision term is identically zero because it is impossible to satisfy 
the delta function constraint. Physically, an on-shell particle of mass 
$m$ cannot decay into two on-shell particles with the same masses. We have 
presented this calculation to illustrate the method that we will use in 
the analysis of more complicated  multi-loop diagrams.

\section{Two-loop contributions}
\label{two-loop}

First we note that at any number of loops greater than one, there are 
two different kinds of diagrams that contribute to the self-energy. The first 
class contains diagrams that are built from diagrams of lower loop order by 
adding tadpole type insertions. An example is shown in Fig.~\ref{tadpole}. 
The contribution to the collision term from a diagram of this type has the 
same structure as the corresponding lower loop diagram with the addition of 
effective masses coming from the tadpole insertions. This type of diagram 
will not be discussed here. The second class of diagrams does not contain 
tadpoles.  This is the type of diagram that we study in this paper. For 
the first few diagrams we give some of the details of the calculation but 
for the most part we focus on the final results.

\par\begin{figure}[H]
\begin{center}
\includegraphics[width=4cm]{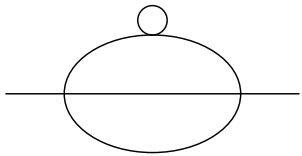}
\end{center}
\caption{Diagram with a tadpole insertion.}
\label{tadpole}
\end{figure}

At the two-loop level, there are two contributions to the self-energy 
in $\phi^3$ theory and one contribution in $\phi^4$ theory (excluding 
diagrams with tadpoles).


\subsection{$\phi^4$ theory}
\label{hh}


The two-loop contribution to self-energy is shown in Fig.~\ref{sun1-B}.
As in the case of the one-loop contribution in $\phi^3$ theory, there 
is only one cut which is the central one. Calculating the retarded, 
advanced, and symmetric components of the self-energy, we obtain
\ba
\label{2-loop}
\Pi^+(p) = \frac{g^2}{24} \int \frac{d^4q}{(2\pi)^4}  
\int \frac{d^4l}{(2\pi)^4} 
&\Big[& \Delta^-(q)\big(\Delta^{\rm sym}(l)\Delta^{\rm sym}(s)
+\Delta^-(l)\Delta^+(s)\big) 
\nonumber\\[2mm]
&+&\Delta^{\rm sym}(q)\big(\Delta^-(l)\Delta^{\rm sym}(s)
+\Delta^{\rm sym}(l)\Delta^+(s)\big) \; \Big] 
\comma
\Pi^-(p) = \frac{g^2}{24} \int \frac{d^4q}{(2\pi)^4}  
\int \frac{d^4l}{(2\pi)^4} &\Big[ & \Delta^{\rm sym}(s)
\big( \Delta^{\rm sym}(q)\Delta^+(l)
+\Delta^{\rm sym}(l)\Delta^+(q)\big) \\[2mm]
&+& \Delta^-(s)\big( \Delta^{\rm sym}(l)\Delta^{\rm sym}(q)
+\Delta^+(l)\Delta^+(q)\big) \; \Big] 
\comma
\Pi^{\rm sym}(p) = \frac{g^2}{24} \int \frac{d^4q}{(2\pi)^4} 
\int \frac{d^4l}{(2\pi)^4} &\Big[& 
\Delta^+(l)\big( \Delta^-(s)\Delta^{\rm sym}(q) 
+ \Delta^{\rm sym}(s)\Delta^+(q)\big)
\nonumber\\[2mm]
&+& \Delta^-(l)\big(\Delta^-(q)\Delta^{\rm sym}(s)
+\Delta^{\rm sym}(q)\Delta^+(s)\big)
\nonumber\\[2mm]
&+& \Delta^{\rm sym}(l)\big(\Delta^{\rm sym}(q)\Delta^{\rm sym}(s)
+\Delta^-(s)\Delta^+(q)+\Delta^-(q)\Delta^+(s)\big) \; \Big] \;,
\ea
where we have used $s \equiv p+q+l$. We substitute these formulas 
into the collision term (\ref{KER}) and express the symmetric Green's
function through the distribution function according to Eq.~(\ref{sym-f0}).
As in the case of the one-loop diagram, we use Eqs.~(\ref{pm0-2},\ref{pole}) 
to find
\ba
\label{ressun1}
\Theta(p_0) \: A(p) \: C_{2L-\phi^4}(p_0) 
&=& \Theta(p_0) \: \frac{g^2 \pi^4}{192} 
\int \frac{d^4q}{(2\pi)^4} \int \frac{d^4l}{(2\pi)^4}
\\ \nonumber 
&\times& \sum_{\{n_p,n_q,n_l,n_s\}=\pm 1} 
{\cal F}_4(n_p,{\bf p}\,;n_q,{\bf q}\,;n_l,{\bf l}\,;-n_s,-{\bf s})
\prod_{\alpha = \{p,\,q,\,l\,s\}} \frac{1}{E_{\alpha}} \: 
\delta(\alpha_0-n_\alpha E_\alpha) \;.
\ea

\par\begin{figure}[H]
\begin{center}
\includegraphics[width=5cm]{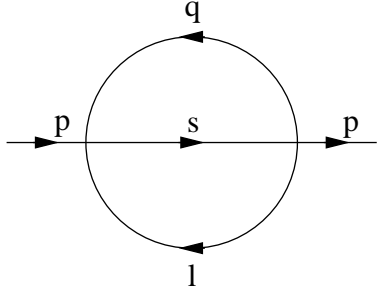}
\end{center}
\caption{Two-loop contribution to the $\phi^4$ self-energy.}
\label{sun1-B}
\end{figure}

We rewrite the result (\ref{ressun1}) in a more symmetric way by 
introducing the delta function $\delta^{(4)}(p+q+l-s)$ and an integral 
over $s$, which allows us to treat $s$ as an independent variable. 
As in the previous section, we use the free spectral function 
(\ref{pos-only}) and consider only positive energy contributions 
to the collision term. Performing the integrals over $q_0$, $l_0$, 
and $s_0$, we have
\ba
\label{h1}
C^{(+)}_{2L-\phi^4}(p) &=&  \frac{g^2 \pi^3}{192}\frac{1}{(2\pi)^8} 
\int d^3 q\int d^3 l\int d^3 s \;
\delta^{(3)}({\bf p}+{\bf q}-{\bf l}-{\bf s}) 
\frac{1}{E_q E_l E_s} \\[2mm]
&\times& \sum_{\{n_p,n_l,n_s\}=\pm 1}
\delta(E_p + n_q E_q + n_l E_l-n_s E_s) \;
{\cal F}_4(1,{\bf p}\,;n_q,{\bf q}\,;n_l,{\bf l}\,;-n_s,-{\bf s})\nonumber
\ea
In principle the sum produces eight terms, but five of these eight terms 
are zero because the corresponding energy conserving delta function has 
no support. There are three choices of $\{n_q,n_l,n_s\}$ that give non-zero 
contributions to the collision integral. These three terms can be combined 
by making an appropriate variable transformation on each term. We list below: 
the three choices for the $n_i$ variables with $i=\{q,\,l,\,s\}$ that give 
non-zero contributions, the variable transformations for each of these 
three terms that allow us to combine them, and the corresponding 
transformations to on-shell four-momenta (using the notation 
(\ref{on-shell})):
\ba
\begin{array}{lccc}
[1] & n_q = - n_l =  n_s = 1; &
{\bf l} \rightarrow -{\bf l}; &
q \rightarrow \tilde q ,~ l\rightarrow -\tilde l,
~s\rightarrow \tilde s ;
\\ [1mm] \label{h2}
[2] & n_q = - n_l = - n_s = - 1 ; &
{\bf q}\rightarrow -{\bf l},~{\bf l}\rightarrow {\bf q} ; &
q \rightarrow -\tilde l,~l\rightarrow \tilde q,
~ s\rightarrow \tilde s ;
\\ [1mm]
[3]& n_q = n_l = n_s = -1; &
{\bf q}\rightarrow -{\bf s},
~{\bf l}\rightarrow-{\bf l},
~{\bf s}\rightarrow -{\bf q} ; &
~q\rightarrow -\tilde s,
~l\rightarrow -\tilde l,
~s\rightarrow -\tilde q .
\end{array}
\ea
The statistical factor and the two delta functions have the same form for 
each term. Combining the deltas into one four-dimensional delta function 
that expresses energy-momentum conservation and using the explicit
form of the statistical factor defined by Eqs.~(\ref{curleyF},\ref{omega}), 
we finally find 
\ba
\label{1st2to2}
C^{(+)}_{2L-\phi^4} [p+q\leftrightarrow l+s]
&=&  \frac{1}{2!} \int \frac{d^3 q}{(2\pi)^3E_q}
\int \frac{d^3 l}{(2\pi)^3E_l}
\int \frac{d^3 s}{(2\pi)^3E_s} \: 
(2\pi)^4\delta^{(4)}(\tilde q+\tilde p-\tilde l-\tilde s) \:
|{\cal M}|^2_{2L-\phi^4} 
\\[2mm] \nonumber
&\times&
\Big[ \big(1+f({\bf p}) \big) \big(1+f({\bf q})\big)
f({\bf l}) \: f({\bf s})
- f({\bf p}) \: f({\bf q}) \big(1+f({\bf l})\big)
\big(1+f({\bf s})\big) \Big] \;,
\ea
where $|{\cal M}|^2_{2L-\phi^4} = g^2/16$. Eq.~(\ref{1st2to2}) 
gives the standard collision term corresponding to the binary 
process symbolically denoted as $p+q\leftrightarrow l+s$ and 
illustrated in Fig. \ref{cross}. The first (second) contribution 
is the usual gain (loss) term.

\par\begin{figure}[H]
\begin{center}
\includegraphics[width=2cm]{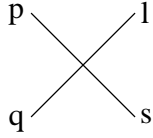}
\end{center}
\caption{$p+q\leftrightarrow l+s$ scattering process.}
\label{cross}
\end{figure}

\subsection{$\phi^3$ theory}
\label{phi3-2L}

In $\phi^3$ theory, there are two graphs of different topologies that 
contribute at two-loop order as shown in Fig. \ref{2loop-topB}.

\par\begin{figure}[H]
\begin{center}
\includegraphics[width=10cm]{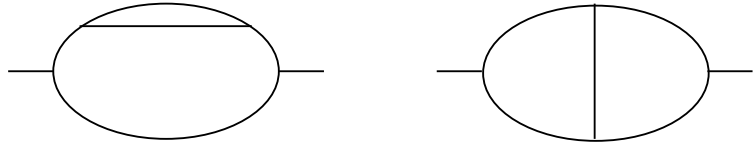}
\end{center}
\caption{Contributions to the self-energy in $\phi^3$ theory.}
\label{2loop-topB}
\end{figure}

We begin with a detailed calculation of the contribution to the 
collision term from the first diagram in Fig. \ref{2loop-SE}.
The results for the self-energies are as follows
\ba
\Pi^+(p) = \frac{g^4}{8}\int \frac{d^4q}{(2\pi)^4}
\int \frac{d^4k}{(2\pi)^4} 
&\Big[& \Delta^-(q)\Delta^-(r)\Delta^{\rm sym}(r)
\big( \Delta^-(l)\Delta^{\rm sym}(k)+\Delta^{\rm sym}(l)\Delta^+(k) \big)
\\
&+&\Delta^-(q)\Delta^-(r)\Delta^+(r) 
\big(\Delta^{\rm sym}(k)\Delta^{\rm sym}(l)
+\Delta^-(l)\Delta^+(k)+\Delta^-(k)\Delta^+(l) \big)
\nonumber \\
&+&\Delta^-(q)\Delta^{\rm sym}(r)\Delta^+(r)
\big( \Delta^-(k)\Delta^{\rm sym}(l)
+\Delta^{\rm sym}(k)\Delta^+(l)\big)
\nonumber \\
&+&\Delta^{\rm sym}(q)\Delta^+(r)^2
\big(\Delta^-(k)\Delta^{\rm sym}(l)
+\Delta^{\rm sym}(k)\Delta^+(l) \big) \; \Big] \;, 
\nonumber\\ [2mm]
\Pi^-(p) = \frac{g^4}{8}\int \frac{d^4q}{(2\pi)^4}
\int \frac{d^4k}{(2\pi)^4}
&\Big[ & \big(\Delta^-(r) \big)^2\Delta^{\rm sym}(q)
\big(\Delta^-(l)\Delta^{\rm sym}(k)+\Delta^{\rm sym}(l)\Delta^+(k) \big)
\\
&+&  \Delta^+(r)\Delta^-(r)\Delta^{\rm sym}(r)
\big(\Delta^-(l)\Delta^{\rm sym}(k)+\Delta^{\rm sym}(l)\Delta^+(k)\big)
\nonumber \\
&+& \Delta^-(r)\Delta^+(p)\Delta^+(r) 
\big(\Delta^{\rm sym}(k)\Delta^{\rm sym}(l)
+\Delta^-(l)\Delta^+(k)+\Delta^-(k)\Delta^+(l) \big)
\nonumber \\
&+& \Delta^{\rm sym}(r)\Delta^+(p)\Delta^+(r)
\big( \Delta^-(k)\Delta^{\rm sym}(l)
+\Delta^{\rm sym}(k)\Delta^+(l) \big) \; \Big] \;, 
\nonumber \\ [2mm]
\Pi^{\rm sym}(p) = \frac{g^4}{8}\int \frac{d^4q}{(2\pi)^4}
\int \frac{d^4k}{(2\pi)^4}
&\Big[& \Delta^-(r)\Delta^{\rm sym}(r)\Delta^{\rm sym}(q)
\big( \Delta^-(l)\Delta^{\rm sym}(k)
+\Delta^{\rm sym}(l)\Delta^+(k) \big)
\\
&+& \Delta^+(r)\Delta^{\rm sym}(r)\Delta^{\rm sym}(q)
\big(\Delta^-(k)\Delta^{\rm sym}(l)
+\Delta^{\rm sym}(k)\Delta^+(l) \big)
\nonumber \\
&+& \big( \Delta^-(r) \big)^2 \Delta^+(q)
\big(\Delta^-(l)\Delta^{\rm sym}(k)
+\Delta^{\rm sym}(l)\Delta^+(k) \big)
\nonumber \\
&+& \big(\Delta^+(r) \big)^2 \Delta^-(q)
\big(\Delta^-(k)\Delta^{\rm sym}(l)
+\Delta^{\rm sym}(k)\Delta^+(l) \big)
\nonumber \\ \nonumber
&+& \Delta^-(r)\Delta^+(r)\Delta^{\rm sym}(q)
\big(\Delta^{\rm sym}(k)\Delta^{\rm sym}(l)
+\Delta^-(l)\Delta^+(k)+\Delta^-(k)\Delta^+(l) \big)\; \Big] \;.
\ea
We substitute these expressions into the collision term (\ref{KER}) and divide 
the results into contributions of three different types, which correspond 
to contributions to the three cuts shown in Fig.~\ref{horz-cuts}. 

\par\begin{figure}[H]
\begin{center}
\includegraphics[width=10cm]{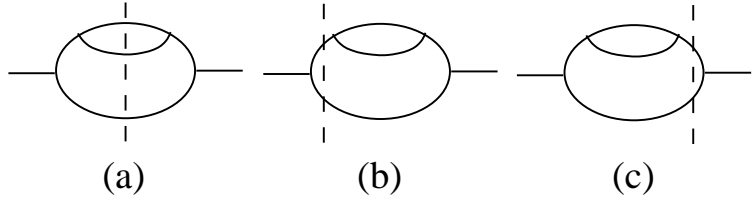}
\end{center}
\caption{Cuts of the first diagram in Fig.~\protect\ref{2loop-topB}.}
\label{horz-cuts}
\end{figure}

\noindent
These cuts are referred to as the `central', `left' and `right' cuts, 
respectively. Terms that contain a factor $\Delta^+(r) \Delta^-(r)$ 
correspond to the central cut. The terms with $\Delta^+(r)$ but without 
$\Delta^-(r)$ are associated with the right cut, and terms that contain 
$\Delta^-(r)$ but not $\Delta^+(r)$ correspond to the left cut. The results 
read:
\ba 
\label{cent-4a}
\Theta(p_0) \: A(p) \: C(p) \Big|_{\rm central} 
&=& \Theta(p_0)\: \frac{g^4\pi^4}{64} 
\int \frac{d^4q}{(2\pi)^4} \int\frac{d^4l}{(2\pi)^4} \: 
\Delta^+(r)\Delta^-(r)  \\ \nonumber
&\times& \sum_{\{n_p,n_q,n_l,n_s\}=\pm 1}
{\cal F}_4(n_p,{\bf p}\,;n_q,{\bf q}\,;n_l,{\bf l}\,;-n_s,-{\bf s})
\prod_{\alpha = \{p,\, q,\, l,\, s\}}\frac{1}{E_\alpha} \: 
\delta(\alpha_0-n_\alpha E_\alpha) \;, \\[3mm] 
\label{right-4a}
\Theta(p_0) \: A(p) \: C(p) \Big|_{\rm right} 
&=& \Theta(p_0)\: \frac{g^2 \pi^3}{32}
\int \frac{d^4q}{(2\pi)^4} \;
\Delta^+(r)\Pi^+(r) \\ \nonumber 
&\times& \sum_{\{n_p,n_q,n_r\}=\pm 1}
{\cal F}_3(n_p,{\bf p}\,;n_q,{\bf q}\,;-n_r,-{\bf r})
\prod_{\alpha = \{p,\, q, \, r\}}\frac{1}{E_\alpha} \:
\delta(\alpha_0-n_\alpha E_\alpha) \;, \\[3mm] 
\label{left-4a}
\Theta(p_0) \: A(p) \: C(p) \Big|_{\rm left} 
&=& \Theta(p_0) \: \frac{g^2 \pi^3}{32}
\int \frac{d^4q}{(2\pi)^4} \;
\Delta^-(r)\Pi^-(r) \\ \nonumber
&\times& \sum_{\{n_p,n_q,n_r\}=\pm 1}
{\cal F}_3(n_p,{\bf p}\,;n_q,{\bf q}\,;-n_r,-{\bf r})
\prod_{\alpha = \{p,\,q,\,r\}}\frac{1}{E_\alpha} \:
\delta(\alpha_0-n_\alpha E_\alpha) \;.
\ea
In Eqs.~(\ref{right-4a}) and (\ref{left-4a}) the factors $\Pi^+(r)$ and 
$\Pi^-(r)$ refer to one-loop self-energies of the form shown in 
Fig.~\ref{fig-one-loop}. Thus, we find that the non-central cuts give 
contributions that can be understood as corrections to the one-loop result 
as shown in Fig.~\ref{Cor-Prop}. 

\par\begin{figure}[H]
\begin{center}
\includegraphics[width=9cm]{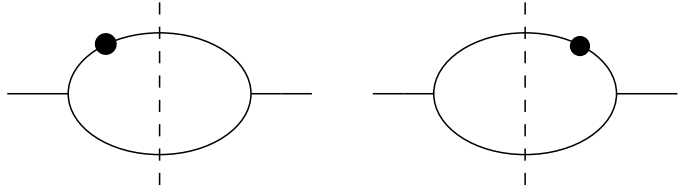}
\end{center}
\caption{Non-central cut for the first diagram in Fig.~\ref{2loop-topB}.}
\label{Cor-Prop}
\end{figure}

Non-central cuts can always be treated in this manner {\it i.e.} every  
non-central cut provides a higher order correction to the central cut of 
some other diagram. As explained in the Introduction, we are only interested 
in the dominant contributions to any given physical process, and consequently 
we consider only central cuts from now on. We calculate contributions to the 
collision term at leading order in $g$ for each physical process from the 
relevant sum of centrally cut self-energy diagrams at the same order of $g$. 
The collision term contains some phase space integrals, the statistical 
factor ${\cal F_N}$ (Eq.~(\ref{curleyF})), and the square of a matrix 
element which is constructed from uncut propagators $\Delta^\pm$. At higher 
orders in $g$, there are many self-energy diagrams that have to be included, 
some of which have more than one central cut. It is a non-trivial technical 
problem to combine all of these terms into a result which can be written 
as the square of a matrix element. This structure is obtained by an 
appropriate choice of momentum labels for the internal lines of the 
self-energy diagrams. We proceed as follows:

\begin{enumerate}

\item[{\bf 1)}] the same set 
of momentum variables is always assigned to the cut propagators  
(equivalently, each central cut has the same ${\cal F}$ factor);

\item[{\bf 2)}] all possible permutations of momentum variables for 
the uncut lines are considered;

\item[{\bf 3)}] for each diagram, all permutations are summed 
and the sum is normalized with the appropriate weight.

\end{enumerate}

Using this strategy, we calculate the nine diagrams shown in 
Figs.~\ref{2loop-SE} and \ref{2loop-Vert} where we have made the 
following definitions for momentum variables:
\ba
\label{mom-def1}
r \equiv p+q \;,\;\;\;\;\;\;
t \equiv q+l \;,\;\;\;\;\;\;
h \equiv p+l \;,\;\;\;\;\;\;
s \equiv r+l\;.
\ea 
We note that the second topology in Fig. \ref{2loop-topB} 
appears twice as often as the first topology (six times as compared 
with three) because of the fact that we take only one of two possible diagonal 
cuts for these diagrams. 

\par\begin{figure}[H]
\begin{center}
\includegraphics[width=10cm]{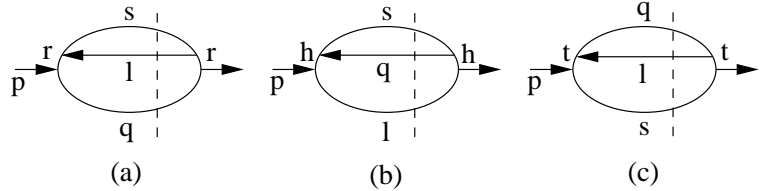}
\end{center}
\caption{Three permutations of the centrally cut first diagram 
in Fig.~\ref{2loop-topB}.}
\label{2loop-SE}
\end{figure}
\par\begin{figure}[H]
\begin{center}
\includegraphics[width=10cm]{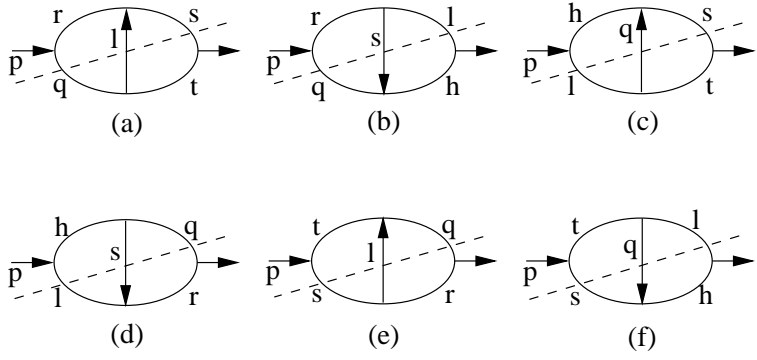}
\end{center}
\caption{Six permutations of the centrally cut second diagram
in Fig.~\ref{2loop-topB}.}
\label{2loop-Vert}
\end{figure}

Defining the operation 
\ba
\label{prefactor}
\ll \cdots \gg_{2L-\phi^3}
&\buildrel \rm def \over = &
\Theta(p_0) \: \frac{g^4\pi^4}{64}\int \frac{d^4q}{(2\pi)^4} 
\int\frac{d^4l}{(2\pi)^4} \\ [2mm]
& \times & \sum_{\{n_p,n_q,n_l,n_s\}=\pm 1}
{\cal F}_4(n_p,{\bf p}\,;n_q,{\bf q}\,;n_l,{\bf l}\,;-n_s,-{\bf s})
\prod_{\alpha = \{p,q,l,s\}}\frac{1}{E_\alpha} \: 
\delta(\alpha_0-n_\alpha E_\alpha) \; \cdots \;,
\nonumber
\ea
the contributions corresponding to the diagrams from 
Figs. \ref{2loop-SE}a-\ref{2loop-Vert}f can be written as
\ba
\begin{array}{ccc}
(\ref{2loop-SE}a) = \;\ll \Delta^+(r)\Delta^-(r) \gg_{2L-\phi^3} \;,&
(\ref{2loop-SE}b) = \;\ll \Delta^+(h)\Delta^-(h) \gg_{2L-\phi^3} \;,&
(\ref{2loop-SE}c) = \;\ll \Delta^+(t)\Delta^-(t) \gg_{2L-\phi^3} \;,
\\[1mm] \nonumber
(\ref{2loop-Vert}a) = \;\ll \Delta^+(r)\Delta^+(t) \gg_{2L-\phi^3} \;,&
(\ref{2loop-Vert}b) = \;\ll \Delta^+(r)\Delta^-(h) \gg_{2L-\phi^3} \;,&
(\ref{2loop-Vert}c) = \;\ll \Delta^+(h)\Delta^+(t) \gg_{2L-\phi^3} \;,
\\[1mm] \nonumber
(\ref{2loop-Vert}d) = \;\ll \Delta^+(h)\Delta^-(r) \gg_{2L-\phi^3} \;,&
(\ref{2loop-Vert}e) = \;\ll \Delta^-(t)\Delta^-(r) \gg_{2L-\phi^3} \;,&
(\ref{2loop-Vert}f) = \;\ll \Delta^-(t)\Delta^-(h) \gg_{2L-\phi^3} .
\end{array}
\ea
After summing these contributions we obtain, 
\ba
\label{ME}
\Theta(p_0) \: A(p) \: C_{2L-\phi^3}(p) = \frac{1}{3} 
\ll \big|\Delta^+(r)+\Delta^+(h)+\Delta^-(t)\big|^2 \gg_{2L-\phi^3} .
\ea

In order to express the result (\ref{ME}) in a more symmetric 
way, we introduce the delta function $\delta^{(4)}(p+q+l-s)$ and an 
integration over $s$, which allows us to treat $s$ as an independent 
variable. As we have done previously, we use Eq.~(\ref{pos-only}) and 
consider only positive energy contributions to the collision term.  
Performing the integrals over $p_0$, $l_0$ and $s_0$, we obtain
\ba
\label{h3}
C^{(+)}_{2L-\phi^3}(p) &=& \frac{g^4 \pi^3}{192}\frac{1}{(2\pi)^8}
\int d^3 q\int d^3 l\int d^3 s ~ 
\delta^{(3)}({\bf q} +{\bf p}+{\bf l}-{\bf s}) 
\: \frac{1}{E_q E_l E_s} 
\\[2mm] \nonumber
&\times& \Big|\Delta^+(p+q)+\Delta^+(p+l)+\Delta^+(-q-l)
\Big|^2_{p_0 = E_q,\, q_0 = n_q E_q, \, l_0 = n_l E_l}
\\[2mm] \nonumber
&\times& \sum_{\{n_q,n_l,n_s\}=\pm 1} 
\delta(E_q +n_p E_p +n_l E_l - n_s E_s )~
{\cal F}_4(1,{\bf p}\,;n_q,{\bf q}\,;n_l,{\bf l}\,;-n_s,-{\bf s}) \;.
\ea
Except for the matrix element, the right-hand-side of this expression 
is the same as the right-hand-side of Eq.~(\ref{h1}). Eqs.~(\ref{h1},\ref{h3}) 
have similar structure since both correspond to contributions to the 
collision term from binary processes. As in the case of Eq.~(\ref{h1}), 
the sum produces three non-zero contributions.  The three choices of the 
$n$'s and the variable transformations that allow us to combine terms are 
the same as in the $\phi^4$ case and are given in Eq.~(\ref{h2}). The 
delta functions and statistical factors are the same as well. A straightforward 
but tedious calculation shows that the matrix element has the same form for 
each term. Thus, we finally obtain 
\ba
\label{2L-phi3-F}
C^{(+)}_{2L-\phi^3} [p+q\leftrightarrow l+s]
&=&  \frac{1}{2!} \int \frac{d^3 q}{(2\pi)^3E_q}
\int \frac{d^3 l}{(2\pi)^3E_l}
\int \frac{d^3 s}{(2\pi)^3E_s} \: 
(2\pi)^4\delta^{(4)}(\tilde q+\tilde p-\tilde l-\tilde s) \:
|{\cal M}|^2_{2L-\phi^3}  
\\[2mm] \nonumber
&\times&
\Big[ \big(1+f({\bf p}) \big) \big(1+f({\bf q})\big)
f({\bf l}) \: f({\bf s})
- f({\bf p}) \: f({\bf q}) \big(1+f({\bf l})\big)
\big(1+f({\bf s})\big) \Big] \;.
\ea
The matrix element squared is given by 
\ba
\big|{\cal M}\big|^2_{2L-\phi^3}  = \frac{g^4}{16} \:
\Big|\frac{1}{S}+\frac{1}{T} +\frac{1}{U}\Big|^2 \;,
\ea
where $S$, $T$, and $U$ are the Mandelstam variables defined as
\ba
S \equiv (\tilde q+\tilde p)^2\;, ~~~~~
T \equiv (\tilde l-\tilde p)^2\;, ~~~~~
U \equiv (\tilde q-\tilde l)^2 \;.
\ea
We have used here capital letters to avoid confusion with momentum 
variables. This matrix element is illustrated in Fig. \ref{2to2new}.

\par\begin{figure}[H]
\begin{center}
\includegraphics[width=5cm]{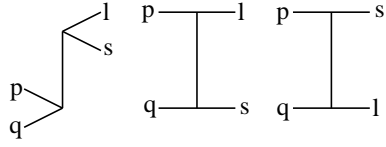}
\end{center}
\caption{Scattering diagrams that contribute to the process 
$p+q\leftrightarrow l+s$ in $\phi^3$ theory.} 
\label{2to2new}
\end{figure}

\section{Three-loop contributions}
\label{three-loop}

\subsection{$\phi^4$ theory}

In $\phi^4$ theory, the only three-loop diagram that does not involve 
tadpoles is shown in Fig.~\ref{3loop-phi4}a. This diagram does not have 
a central cut. The non-central cut presented in Fig.~\ref{3loop-phi4}a 
produces a correction to the two-loop sunset graph, as shown in 
Fig.~\ref{3loop-phi4}b, and will not be discussed.

\par\begin{figure}[H]
\begin{center}
\includegraphics[width=6cm]{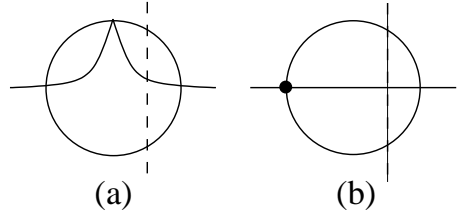}
\end{center}
\caption{Three-loop contribution to the $\phi^4$ self-energy (a);
and an effective two-loop contribution containing a corrected
vertex function (b).}
\label{3loop-phi4}
\end{figure}

\subsection{$\phi^3$ theory}

There are diagrams of eight different topologies which contribute
to the $\phi^3$ self-energy at the three-loop level. They are shown in 
Fig.~\ref{3L-phi3}.

\par\begin{figure}[H]
\begin{center}
\includegraphics[width=8.5cm]{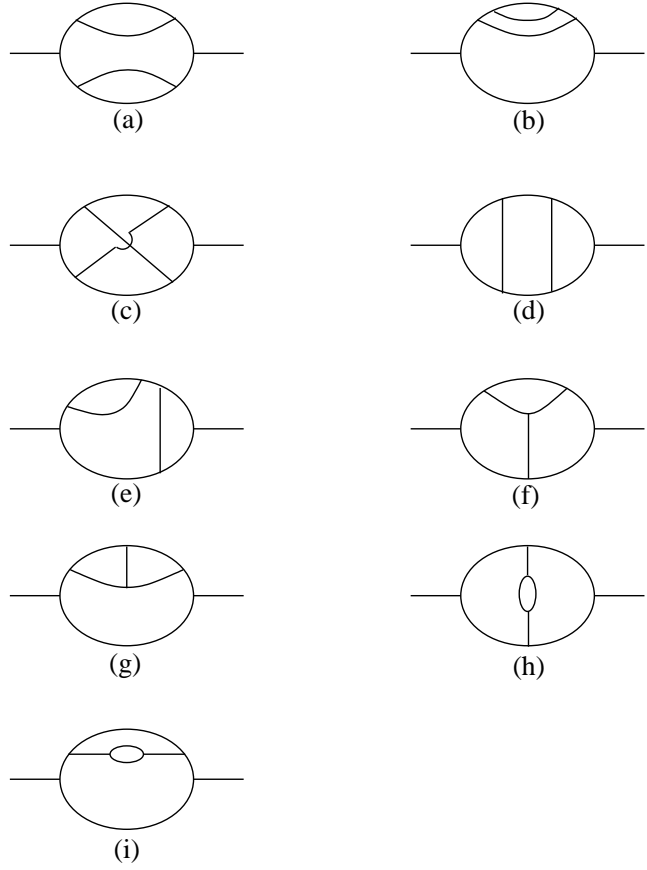}
\end{center}
\caption{Three-loop contributions to the $\phi^3$ self-energy.}
\label{3L-phi3}
\end{figure}

We follow the same strategy as in the previous section and consider
only the central cuts of the diagrams in Fig.~\ref{3L-phi3}. The 
internal lines are labeled in all possible ways that produce the same  
statistical factor. In addition to the definition (\ref{mom-def1}),
we define the following momentum variables:
\ba
\label{mom-def2}
y \equiv q + k        \;,~~~~~
m \equiv k + l        \;,~~~~~
j \equiv k + p        \;,~~~~~
z \equiv p + q + k    \;, \\[2mm] \nonumber
w \equiv q + k + l    \;,~~~~~
v \equiv p + k + l    \;,~~~~~
u \equiv p + q + k +l \;. 
\ea
In analogy to Eq.~(\ref{prefactor}), we define the operation  
\ba
\ll \cdots \gg_{3L-\phi^3} 
&\buildrel \rm def \over =&
\Theta(p_0) \: \frac{g^6\pi^5}{64}
\int\frac{d^4k}{(2\pi)^4}
\int\frac{d^4l}{(2\pi)^4}
\int\frac{d^4q}{(2\pi)^4} 
\\[2mm] \nonumber
&\times& \sum_{\{n_p,n_q,n_l,n_k,n_u\}=\pm 1}
{\cal F}_5(n_p,{\bf p}\,;n_q,{\bf q}\,;n_l,{\bf l}\,;
n_k,{\bf k}\,;-n_u,-{\bf u})
\prod_{\alpha = \{p,\,q,\,l,\,k,u\}} \frac{1}{E_\alpha} \:
\delta(\alpha_0-n_\alpha E_\alpha) ~~\cdots
\ea
Combining the results for the diagrams in 
Figs.~\ref{3L-phi3}a-~\ref{3L-phi3}h, we obtain
\ba
\label{ME-3L}
\Theta(p_0) \: A(p) \: C_{3L-\phi^3}(p) 
&=& \frac{1}{24} \ll \Big|
   \Delta^+(t)\Delta^-(s)
+  \Delta^+(y)\Delta^-(z)
+  \Delta^+(m)\Delta^-(v)  
\\[2mm] \nonumber
&+&\Delta^+(m)\Delta^-(r)
+  \Delta^-(r)\Delta^-(z)
+  \Delta^-(r)\Delta^-(s)
+  \Delta^-(h)\Delta^-(v)
\\[2mm] \nonumber
&+& \Delta^+(y)\Delta^-(h)
+  \Delta^-(h)\Delta^-(s)
+  \Delta^+(t)\Delta^-(j)
+  \Delta^-(v)\Delta^-(j)
\\[2mm] \nonumber
&+&\Delta^-(z)\Delta^-(j)
+  \Delta^+(m)\Delta^+(w)
+  \Delta^+(t)\Delta^+(w)
+  \Delta^+(y)\Delta^+(w) \Big|^2 \gg_{3L-\phi^3} \;.
\ea
We can rewrite the result (\ref{ME-3L}) by introducing the delta function 
$\delta^{(4)}(p+q+k+l-u)$ and an integral over $u$. As we have done 
previously, we use Eq.~(\ref{pos-only}) and consider only positive energy 
contributions to the collision term. The sum over $n_q,~n_l,~n_k$, and 
$n_u$ produces 16 terms. Of these 16 terms, 10 are kinematically allowed. 
Four of these processes can be combined by relabeling momentum variables 
to produce the process symbolically denoted as  
$p+q \leftrightarrow l+k+u$. The remaining six can be combined 
to produce the process of the form 
$p+q+l\leftrightarrow k+u$. \\

\underline{\sc The process $p + q \leftrightarrow l + k + u$}\\

We list the four possible choices of the $n$'s and the variable 
transformations that allow us to combine terms and obtain the matrix 
element corresponding to the process $p + q \leftrightarrow l + k + u$:
\ba
\begin{array}{lccc}
[1] & n_q =- n_l = - n_k = n_u = 1;&
{\bf l} \rightarrow - {\bf l},~ 
{\bf k} \rightarrow - {\bf k} ;&
q \rightarrow  \tilde q,~
l \rightarrow -\tilde l,
\\ & & &
k \rightarrow -\tilde k,~
u \rightarrow  \tilde u ; 
\\[3mm]
[2] & n_q = n_l = - n_k = - n_u = - 1;&
{\bf q} \rightarrow -{\bf k},~
{\bf k} \rightarrow  {\bf q}, &
q\rightarrow -\tilde k,~l\rightarrow -\tilde l,
\\ & & 
{\bf l} \rightarrow -{\bf l} ;& 
k\rightarrow \tilde q,~u\rightarrow \tilde u;
\\[3mm]
[3] & n_q = - n_l = n_k = - n_u = -1; &
{\bf q} \rightarrow -{\bf l},~
{\bf l} \rightarrow  {\bf q},~&
q \rightarrow -\tilde l,~
l \rightarrow  \tilde q,
\\ & & 
{\bf k} \rightarrow -{\bf k} ;&
k \rightarrow -\tilde k,~
u \rightarrow  \tilde u ;
\\[3mm]
[4] & n_q = n_l =  n_k = n_u = -1;&
{\bf q} \rightarrow -{\bf u},~
{\bf l} \rightarrow -{\bf l},&
q \rightarrow -\tilde u,~
k \rightarrow -\tilde k,
\\ & & 
{\bf k} \rightarrow -{\bf k},~
{\bf u} \rightarrow -{\bf q} ;&
l \rightarrow -\tilde l,~
u \rightarrow -\tilde q .
\end{array}
\ea
The delta functions and statistical factor have the same form for each 
of these four terms. A straightforward but very tedious calculation 
verifies that the same matrix element is produced in each case. The 
collision term representing the process $p+q\leftrightarrow l+k+u$
equals
\ba
\label{2to3}
C^{(+)}_{3L-\phi^3}
[p+q\leftrightarrow l+k+u] 
&=& \frac{1}{3!}
\int \frac{d^3q}{(2\pi)^3E_q} 
\int \frac{d^3l}{(2\pi)^3E_l}
\int \frac{d^3k}{(2\pi)^3E_k}
\int \frac{d^3u}{(2\pi)^3E_u}
\\[2mm] \nonumber
&\times& 
(2\pi)^4 \delta^{(4)}(\tilde p+\tilde q-\tilde l-\tilde k-\tilde u)~
|{\cal M}|^2_{3L-\phi^3}[p+q\leftrightarrow l+k+u] 
\\[2mm] \nonumber
&\times& \Big[
\big(1+f({\bf p})\big)\: \big(1+f({\bf q})\big) 
\: f({\bf l}) \: f({\bf k}) \: f({\bf u})
- f({\bf p})\: f({\bf q}) \:\big(1+ f({\bf l})\big) \: 
\big(1+f({\bf k}) \big) \: \big(1+ f({\bf u})\big) \Big] \;, 
\ea
where the matrix element squared of the diagrams shown in 
Fig.~\ref{2to3new} is
\ba
\label{2to3-matrix}
|{\cal M}|^2_{3L-\phi^3}[p+q\leftrightarrow l+k+u] 
& = &\frac{g^6}{2^5} \Big|
\:\Delta^-(\tilde p+\tilde q) \big[
 \Delta^-(\tilde k+\tilde l)+\Delta^-(\tilde k+\tilde u)
+\Delta^-(\tilde l+\tilde u)\big]
\\ [2mm] \nonumber 
&+&~~\Delta^-(\tilde p-\tilde l) \big[
 \Delta^+(\tilde q-\tilde k)+\Delta^+(\tilde q-\tilde u)
+\Delta^-(\tilde k+\tilde u)\big] 
\\ [2mm] \nonumber
&+&~~\Delta^-(\tilde p-\tilde k)\big[
 \Delta^+(\tilde q-\tilde l) + \Delta^+(\tilde q-\tilde u)
+\Delta^-(\tilde l+\tilde u) \big]
\\ [2mm] \nonumber
&+&~~\Delta^-(\tilde p-\tilde u)\big[\Delta^+(\tilde q-\tilde l)
 + \Delta^+(\tilde q-\tilde k)+\Delta^-(\tilde l+\tilde k)\big]
\\ [2mm] \nonumber
&+& 
 \Delta^+(\tilde q-\tilde l)\Delta^-(\tilde k+\tilde u) 
+\Delta^+(\tilde q-\tilde k)\Delta^-(\tilde l+\tilde u)
+\Delta^+(\tilde q-\tilde u)\Delta^-(\tilde l+\tilde k)  \Big|^2 .
\ea

\par\begin{figure}[H]
\begin{center}
\includegraphics[width=10cm]{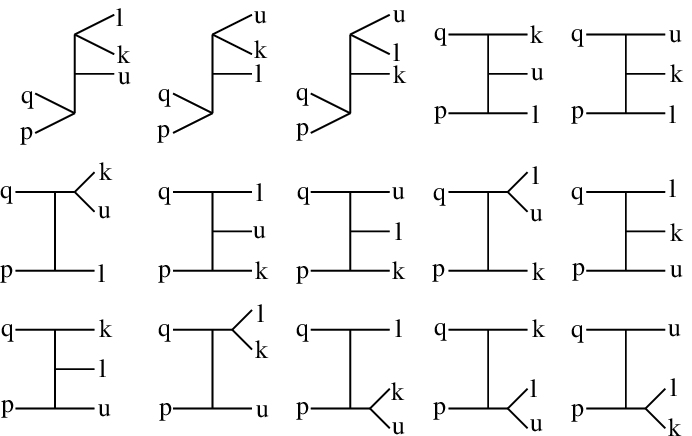}
\end{center}
\caption{Scattering diagrams that contribute to the process 
$p+q\leftrightarrow l+k+u$  in the order they appear in 
Eq.~(\ref{2to3-matrix}).}
\label{2to3new}
\end{figure}

\vspace*{.5cm}

\underline{\sc The process $q+p+l\leftrightarrow k+u$ }\\

We list the six possible choices of the $n$'s and the variable 
transformations that allow us to combine terms and obtain the matrix 
element corresponding to the process $p+q+l\leftrightarrow k+u$:  
\ba
\begin{array}{lccc}
[1] & n_q=n_l=-n_k=~n_u=1; &
{\bf k}\rightarrow -{\bf k}; &
k\rightarrow -\tilde k; 
\\[3mm]
[2] & n_q=-n_l=n_k=n_u=1; &
{\bf l}\ra -{\bf k},~{\bf k}\ra {\bf l};&
l\ra -\tilde k,~k\ra \tilde l; 
\\[3mm]
[3] & n_q=-n_l=-n_k=-n_u=-1;&
{\bf p}\ra -{\bf k},~{\bf k}\ra {\bf p};&
p\ra -\tilde k,~k\ra \tilde p; 
\\[3mm]
[4] & n_q=-n_l=-n_k=-n_u=1;&
{\bf l}\ra -{\bf u},~{\bf k}\ra -{\bf k},&
l\ra -\tilde u,~k\ra -\tilde k,
\\ & &
{\bf u}\ra -{\bf l};&
~u\ra -\tilde l; 
\\[3mm]
[5] & n_q=-n_l=n_k=n_u=-1;&
{\bf p}\ra-{\bf u},~{\bf k}\ra-{\bf l},&
p\ra -\tilde u,~k\ra -\tilde k,
\\ & &
{\bf u}\ra-{\bf p};&
u\ra -\tilde p; 
\\[3mm]
[6] & n_q=n_l=-n_k=n_u=-1;&
{\bf p}\ra -{\bf u},~{\bf l}\ra -{\bf k},&
p\ra -\tilde u,~l\ra -\tilde k,
\\ & &
{\bf k}\ra {\bf l},~{\bf u}\ra -{\bf p};&
k\ra \tilde l,~u\ra -\tilde p.
\end{array}
\ea
The delta functions and statistical factors have the same form for each 
of these six terms. Similarly, one can verify that the matrix element is 
the same in each case. The contribution to the collision term from  
$p+q+l\leftrightarrow k+u$ processes is
\ba
\label{3to2}
C^{(+)}_{3L-\phi^3}[p+q+l \leftrightarrow k+u]
&=& 
\frac{1}{2! \: 2!}
\int \frac{d^3q}{(2\pi)^3E_q}
\int \frac{d^3l}{(2\pi)^3E_l}
\int \frac{d^3k}{(2\pi)^3E_k}
\int \frac{d^3u}{(2\pi)^3E_u}
\\[2mm] \nonumber
&\times&
(2\pi)^4 \delta^{(4)}(\tilde p+\tilde q-\tilde l-\tilde k-\tilde u)~
|{\cal M}|^2_{3L-\phi^3}[p+q+l \leftrightarrow k+u]
\\[2mm] \nonumber
&\times& \Big[
\big(1+f({\bf p})\big)\: \big(1+f({\bf q})\big)
\: \big(1+f({\bf l})\big) \: f({\bf k}) \: f({\bf u})
- f({\bf p})\: f({\bf q}) \: f({\bf l})\big) \:
\big(1+f({\bf k}) \big) \: \big(1+ f({\bf u})\big) \Big] \;,
\ea
where the matrix element squared of the diagrams shown in
Fig.~\ref{3to2new} equals
\ba
\label{3to2-matrix}
|{\cal M}|^2_{3L-\phi^3}[p+q+l \leftrightarrow k+u]
& = &\frac{g^6}{2^5} \Big|
\Delta^-(\tilde p+\tilde q)
\big[\Delta^-(\tilde k-\tilde l)+\Delta^-(\tilde k+\tilde u)
+\Delta^-(-\tilde l+\tilde u)\big]
\\[2mm] \nonumber
&+&~~\Delta^-(\tilde p + \tilde l)
\big[\Delta^+(\tilde q-\tilde k) +\Delta^+(\tilde q-\tilde u)
+\Delta^-(\tilde k+\tilde u)\big]
\\[2mm] \nonumber
&+&~~\Delta^-(\tilde p-\tilde k)
\big[\Delta^+(\tilde q+\tilde l)+\Delta^+(\tilde q-\tilde u)
+\Delta^-(-\tilde l+\tilde u)\big]
\\[2mm] \nonumber
&+&~~\Delta^-(\tilde p-\tilde u)
\big[\Delta^+(\tilde q+\tilde l)+\Delta^+(\tilde q-\tilde k)
+\Delta^-(-\tilde l+\tilde k)\big]
\\[2mm] \nonumber
&+&  \Delta^+(\tilde q+\tilde l)\Delta^-(\tilde k+\tilde u) 
         + \Delta^+(\tilde q-\tilde k)\Delta^-(-\tilde l+\tilde u)
         + \Delta^+(\tilde q-\tilde u)\Delta^-(-\tilde l+\tilde k)
 \Big|^2 .
\ea

\par\begin{figure}[H]
\begin{center}
\includegraphics[width=10cm]{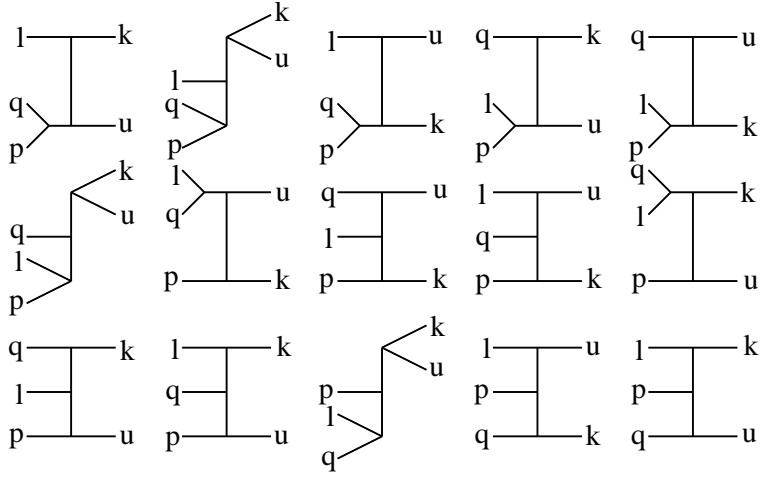}
\end{center}
\caption{Scattering diagrams that contribute to the process 
$p+q+l\leftrightarrow k+u$ in the order they appear in 
Eq.~(\ref{3to2-matrix}).}
\label{3to2new}
\end{figure}

The collision term for $\phi^3$ theory at the three-loop level, 
which represents the $2 \leftrightarrow 3$ processes, is given by the 
sum of the contributions (\ref{2to3},\ref{3to2}). Such a form of the 
collision term was postulated in \cite{Jeon:1994if,Jeon:1995zm} to 
reproduce the Kubo formula result of the viscosity coefficients within 
a linearized kinetic theory.

\section{Four-loop contributions}
\label{four-loop}

At the four-loop level, we consider only $\phi^4$ theory. The relevant diagrams
are the double sunset and two crossed versions of the double sunset, as shown 
in Fig~\ref{sun2-all}. All other contributions contain tadpoles and will not 
be discussed, as has been previously explained.

\par\begin{figure}[H]
\begin{center}
\includegraphics[width=14cm]{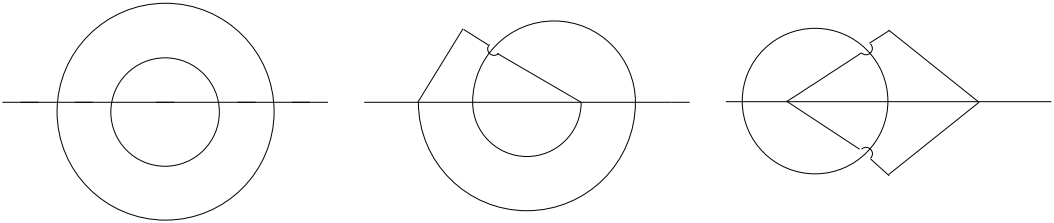}
\end{center}
\caption{Four-loop contributions to the $\phi^4$ self-energy.}
\label{sun2-all}
\end{figure}

We follow the strategy formulated in Sec.~\ref{phi3-2L} and label internal
lines of the diagrams  in all possible ways that produce the same statistical 
factor. As previously, we define the operation
\ba
\ll \cdots \gg_{4L-\phi^4} &\buildrel \rm def \over =& 
\Theta(p_0) \: \frac{g^4\pi^6}{1536}\int \frac{d^4q}{(2\pi)^4}
\int \frac{d^4l}{(2\pi)^4}\int \frac{d^4m}{(2\pi)^4}
\int \frac{d^4k}{(2\pi)^4}
\nonumber \\ [2mm] 
&\times&  \sum_{\{n_p,n_q,n_l,n_m,n_k,n_x\}=\pm1}
{\cal F}_6(n_p,{\bf p}\,;n_q,{\bf q}\,;
n_l,{\bf l}\,;n_m,{\bf m}\,;n_k,{\bf k}\,;-n_x,-{\bf x})
\nonumber \\ [2mm] \label{4TO4}
&\times& \prod_{\alpha = \{p,\,q,\,l,\,m,\,k,\,x\}}
\frac{1}{E_\alpha}\delta(\alpha_0-n_\alpha E_\alpha)~\cdots
\ea
where we have used the definitions (\ref{mom-def1}) and (\ref{mom-def2}), 
and additionally, we have introduced $x \equiv q+p+k+l+m$. After summing up
all central cuts of the diagrams shown in Fig~\ref{sun2-all}, we get the 
four-loop collision term
\ba
\Theta(p_0) \: A(q) \: C_{4L-\phi^4}(p) 
&=& \frac{1}{10}\ll \big|
\Delta^+(p+q+k)+\Delta^+(p+q+l)+\Delta^+(p+q+m)
\\ [2mm] \nonumber
&+&\Delta^+(p+k+l) +\Delta^+(p+k+m)+\Delta^+(p+l+m) 
\\ [2mm] \nonumber
&+&\Delta^+(k+l+m)+\Delta^+(q+k+l)
+\Delta^+(q+l+m)+\Delta^+(q+k+m)\big|^2 \gg_{4L-\phi^4} \;.
\nonumber
\ea
We rewrite this result by introducing the delta function
$\delta^{(4)}(q+p+k+l+m-x)$ and an integral over $x$. As we have 
done previously, we will use Eq.~(\ref{pos-only}) and consider only 
positive energy 
contributions to the collision term. The sum over the remaining $n$'s 
contains 32 terms. Five of these terms can be combined to give the 
process $p+q\leftrightarrow k+l+m+x$, ten give the process  
$p+q+l+m\leftrightarrow k+x$, and ten give the process 
$p+q+l\leftrightarrow m+k+x$. The remaining seven terms are
kinematically forbidden.  Below, we give the collision terms 
representing these processes.\\

\underline{\sc The process $p+q\leftrightarrow k+l+m+x$ }\\

\ba
\label{2to4}
C^{(+)}_{4L-\phi^4}[p+q\leftrightarrow l+k+m+x] 
&=& 
\frac{1}{4!}
\int \frac{d^3q}{(2\pi)^3E_q} 
\int \frac{d^3l}{(2\pi)^3E_l}
\int \frac{d^3k}{(2\pi)^3E_k}
\int \frac{d^3m}{(2\pi)^3E_m}
\int \frac{d^3x}{(2\pi)^3E_x}
\\[2mm] \nonumber
&\times&
(2\pi)^4 \delta^{(4)}(\tilde p+\tilde q-\tilde l-\tilde k-\tilde m - \tilde x)~
|{\cal M}|^2_{4L-\phi^4}[p+q\leftrightarrow l+k+m+x] 
\\[2mm] \nonumber
&\times& \Big[
\big(1+f({\bf p})\big)\: \big(1+f({\bf q})\big) 
\: f({\bf l}) \: f({\bf k}) \: f({\bf m}) \: f({\bf x})
\\[2mm] \nonumber
&-& ~~ f({\bf p})\: f({\bf q}) \:\big(1+ f({\bf l})\big) \: 
\big(1+f({\bf k}) \big) \: \big(1+ f({\bf m})\big) 
\big(1+ f({\bf x})\big) \Big] \;, 
\ea
where the matrix element squared of the diagrams shown in 
Fig.~\ref{2to4new} equals
\ba
\label{2to4-matrix}
|{\cal M}|^2_{4L-\phi^4}[p+q\leftrightarrow l+k+m+x] 
&= &\frac{g^4}{2^6} \Big|
   \Delta^-(\tilde k+\tilde l +\tilde m) 
 + \Delta^+(\tilde q-\tilde k -\tilde l)
 + \Delta^+(\tilde q-\tilde k -\tilde m)
\\[2mm] \nonumber 
&+&\Delta^+(\tilde q-\tilde l -\tilde m) 
 + \Delta^+(\tilde p-\tilde k -\tilde l) 
 + \Delta^+(\tilde p-\tilde k -\tilde m) 
\\[2mm] \nonumber 
&+&\Delta^+(\tilde p-\tilde l -\tilde m) 
 + \Delta^+(\tilde p+\tilde q -\tilde k) 
 + \Delta^+(\tilde p+\tilde q -\tilde l) 
 + \Delta^+(\tilde p+\tilde q -\tilde m)
\Big|^2.
\ea

\par\begin{figure}[H]
\begin{center}
\includegraphics[width=10cm]{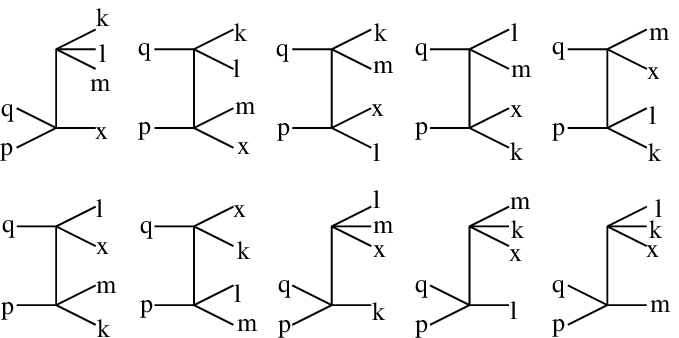}
\end{center}
\caption{Scattering diagrams that contribute to the process 
$p+q\leftrightarrow k+l+m+x$ in the order they appear in 
Eq.~(\ref{2to4-matrix}).}
\label{2to4new}
\end{figure}

\underline{\sc The process $p+q+l+m \leftrightarrow k+x$}\\

\ba
\label{4to2}
C^{(+)}_{4L-\phi^4}
[p+q+l+m\leftrightarrow k+x] 
&=& 
\frac{1}{3! \: 2!}
\int \frac{d^3q}{(2\pi)^3E_q} 
\int \frac{d^3l}{(2\pi)^3E_l}
\int \frac{d^3m}{(2\pi)^3E_m}
\int \frac{d^3k}{(2\pi)^3E_k}
\int \frac{d^3x}{(2\pi)^3E_x}
\\[2mm] \nonumber
&\times&
(2\pi)^4 \delta^{(4)}(\tilde p+\tilde q+\tilde l+\tilde m-\tilde k - \tilde x)~
|{\cal M}|^2_{4L-\phi^4}[p+q+l+m\leftrightarrow k+x] 
\\[2mm] \nonumber
&\times& \Big[
\big(1+f({\bf p})\big)\: \big(1+f({\bf q})\big) \: 
\big(1+ f({\bf l})\big) \: \big(1+ f({\bf m})\big) \: f({\bf k}) \: f({\bf x})
\\[2mm] \nonumber
&-& ~~ f({\bf p})\: f({\bf q}) \:f({\bf l})\big) \: f({\bf m}) \big) \: 
\big(1+ f({\bf k}) \big) \: \big(1+ f({\bf x})\big) \Big] \;, 
\ea
where the matrix element squared of the diagrams shown in 
Fig.~\ref{4to2new} equals
\ba
\label{4to2-matrix}
|{\cal M}|^2_{4L-\phi^4}[p+q+l+m\leftrightarrow k+x] 
 &= &\frac{g^4}{2^6} \Big|
    \Delta^+(\tilde l +\tilde m -\tilde k) 
 +  \Delta^+(\tilde q -\tilde k +\tilde l) 
 +  \Delta^+(\tilde q -\tilde k +\tilde m) 
\\[2mm] \nonumber 
&+& \Delta^+(\tilde q +\tilde l +\tilde m) 
 +  \Delta^+(\tilde p -\tilde k +\tilde l) 
 +  \Delta^+(\tilde p -\tilde k +\tilde m) 
\\[2mm] \nonumber 
&+& \Delta^+(\tilde p +\tilde l +\tilde m) 
 +  \Delta^+(\tilde p +\tilde q -\tilde k) 
 +  \Delta^+(\tilde p +\tilde q +\tilde l) 
 +  \Delta^+(\tilde p +\tilde q +\tilde m)
\Big|^2.
\ea

\par\begin{figure}[H]
\begin{center}
\includegraphics[width=10cm]{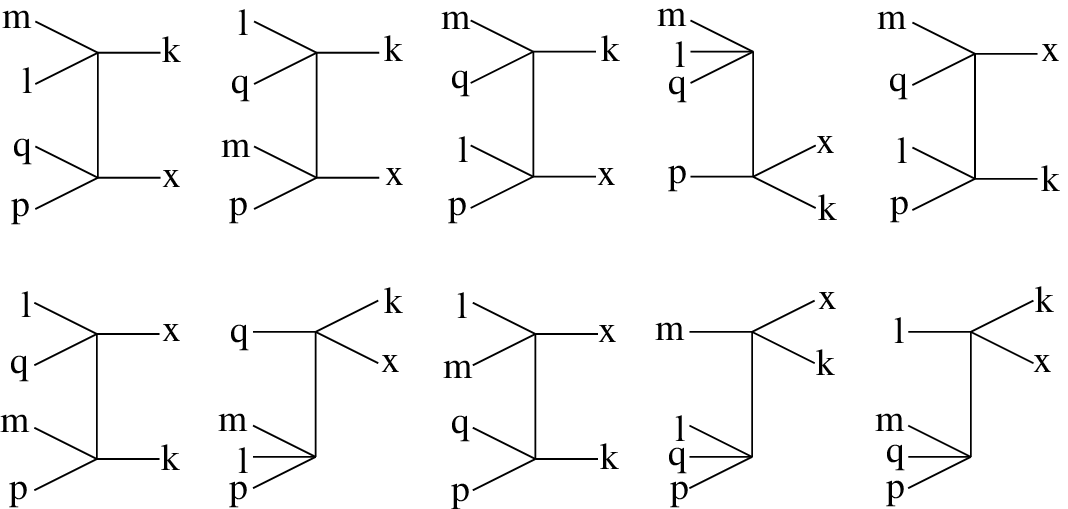}
\end{center}
\caption{Scattering diagrams that contribute to the process 
$p+q+l+m\leftrightarrow k+x$ in the order they appear in 
Eq.~(\ref{4to2-matrix}).}
\label{4to2new}
\end{figure}

\underline{\sc The process $p+q+l \leftrightarrow m+k+x$}\\

\ba
\label{3to3}
C^{(+)}_{4L-\phi^4}
[p+q+l \leftrightarrow m+k+x] 
&=& 
\frac{1}{2! \: 3!}
\int \frac{d^3q}{(2\pi)^3E_q} 
\int \frac{d^3l}{(2\pi)^3E_l}
\int \frac{d^3m}{(2\pi)^3E_m}
\int \frac{d^3k}{(2\pi)^3E_k}
\int \frac{d^3x}{(2\pi)^3E_x}
\\[2mm] \nonumber
&\times&
(2\pi)^4 \delta^{(4)}(\tilde p+\tilde q+\tilde l -\tilde m-\tilde k - \tilde x)~
|{\cal M}|^2_{4L-\phi^4}[p+q+l \leftrightarrow m+k+x] 
\\[2mm] \nonumber
&\times& \Big[
\big(1+f({\bf p})\big)\: \big(1+f({\bf q})\big) \: 
\big(1+ f({\bf l})\big) \:f({\bf m}) \: f({\bf k}) \: f({\bf x})
\\[2mm] \nonumber
&-& ~~ f({\bf p})\: f({\bf q}) \:f({\bf l}) \:\big(1+ f({\bf m}) \big) \: 
\big(1+ f({\bf k}) \big) \: \big(1+ f({\bf x})\big) \Big] \;, 
\ea
where the matrix element squared of the diagrams shown in 
Fig.~\ref{3to3new} equals
\ba
\label{3to3-matrix}
|{\cal M}|^2_{4L-\phi^4}[p+q+l \leftrightarrow m+k+x] 
 &= &\frac{g^4}{2^6} \Big|
   \Delta^+(\tilde l -\tilde k -\tilde m) 
 + \Delta^+(\tilde q -\tilde k +\tilde l) 
 + \Delta^+(\tilde q -\tilde k -\tilde m)
\\[2mm]\nonumber
&+&\Delta^+(\tilde q +\tilde l -\tilde m) 
 + \Delta^+(\tilde p -\tilde k +\tilde l) 
 + \Delta^+(\tilde p -\tilde k -\tilde m) 
\\[2mm]\nonumber
&+&\Delta^+(\tilde p +\tilde l -\tilde m) 
 + \Delta^+(\tilde p +\tilde q -\tilde k) 
 + \Delta^+(\tilde p +\tilde q +\tilde l) 
 + \Delta^+(\tilde p +\tilde q -\tilde m)
\Big|^2 .
\ea

\par\begin{figure}[H]
\begin{center}
\includegraphics[width=10cm]{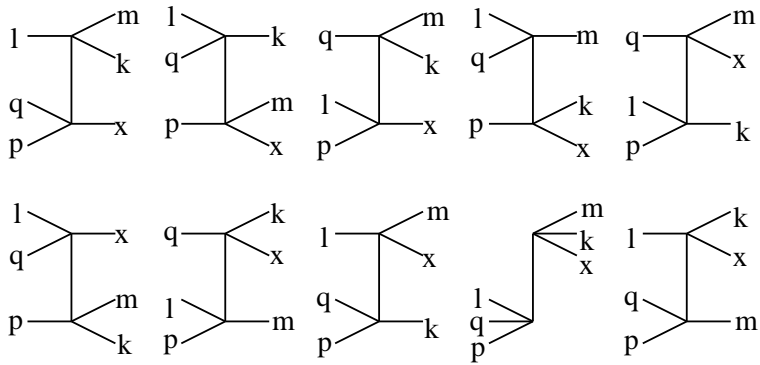}
\end{center}
\caption{Scattering diagrams that contribute to the process 
$p+q+l\leftrightarrow m+k+x$  in the order they appear in 
Eq.~(\ref{3to3-matrix}).}
\label{3to3new}
\end{figure}

The collision term of the $\phi^4$ theory at four-loop level, 
which represents the $2 \leftrightarrow 4$ and $3 \leftrightarrow 3$ 
processes, is given by the sum of the contributions 
(\ref{2to4},\ref{4to2},\ref{3to3}). As in the case of $\phi^3$ interaction, 
a form of the collision term beyond binary approximation was postulated 
in \cite{Jeon:1994if,Jeon:1995zm} to reproduce the Kubo formula result 
of the viscosity coefficients within a linearized kinetic theory. While 
the contributions representing the $2 \leftrightarrow 4$ processes 
coincide, the term corresponding to the $3 \leftrightarrow 3$ interactions 
is absent in the effective transport equation \cite{Jeon:1994if,Jeon:1995zm},
as it is supposed to give negligible contribution to the viscosity 
\cite{Jeon2004}. As already mentioned, the scattering $3 \leftrightarrow 3$ 
was discussed in \cite{Xu:2004gw}.

In performing the perturbative calculations of the collision terms in 
Secs. \ref{one-loop}-\ref{four-loop}, we have used non-interacting Green's 
functions. However, the transport equation is of interest precisely because 
its solution is the full interacting distribution function. The terms on the 
left hand side of the transport equation clearly depend on this 
interacting distribution function. On the right hand side, we use the 
perturbatively calculated collision term and replace the non-interacting 
distribution function by the interacting one. We drop gradient terms in 
the expansion of the interacting distribution function about the 
non-interacting distribution, when these gradient terms appear within 
the collision term. The justification for this procedure is as follows: 
we consider the gradient and coupling constant approximations as independent, 
and choose to investigate higher order terms in the coupling constant 
approximation, while working at lowest order in the gradient approximation. 
The contributions we have calculated correspond physically to multi-particle 
production processes involving zero width quasi-particles.  We remind the 
reader that finite width quasi-particles will give rise to contributions 
to the collision term that are kinematically forbidden for zero-width 
quasi-particles. For $\phi^4$ theory an example is a $1 \leftrightarrow 3$ 
process. In this paper we assume infinitely narrow quasi-particles 
(Eq. (\ref{quasi-particle-approx-2})), and in this limit the effects 
of these processes will be negligible.

\section{Summary and outlook}
\label{discussion}

Kinetic equations usually take into account the interaction of particles with 
a mean field, and inter-particle pair collisions. However, when the system 
of interest is very dense, multi-particle interactions are expected to
play a significant role in the system's dynamics. In addition, if a 
characteristic particle's kinetic energy is comparable to the particle's 
mass, particle production processes become important. Both multi-particle 
interactions and production processes occur in systems of relativistic 
quantum fields, such a quark-gluon plasma, when the energy density is 
sufficiently high. 

In this study, we have given the first systematic derivation of the 
transport equation of relativistic quantum fields which takes into 
account multi-particle and production process. Using the Schwinger-Keldysh 
approach, we have discussed scalar fields with cubic and quartic 
interactions. Mean-field phenomena are controlled by the one-loop 
tadpole contributions to the self-energy, and binary collisions correspond 
to the two-loop graphs. Multi-particle process occur at the three-loop
level in $\phi^3$ theory and at the four-loop level in $\phi^4$ theory.
Analysis of three- and four-loop diagrams is a technically very
complex problem which requires special computational techniques. 
We have used the Keldysh representation because it provides a clean 
separation between propagators that correspond to real and virtual particles. 
In addition, we have made use of the cut structure of the collision term.  
Our calculations have been performed with the help of a program that uses 
{\it MATHEMATICA} symbolic manipulation software. 

Throughout this paper, we have assumed that quasi-particles are of zero-width 
and that their four-momenta are on the mass-shell. This assumption is crucial 
since some processes which occur at one- and two-loop level, like 
$1 \leftrightarrow 2$ and $1 \leftrightarrow 3$ processes, are kinematically 
forbidden for on mass-shell particles. We have derived the explicit form 
of the contributions to the collision term of the transport equation 
corresponding to the processes: $2 \leftrightarrow 3$ for the $\phi^3$ model, 
and $2 \leftrightarrow 4$ and $3 \leftrightarrow 3$ for the $\phi^4$ model.

In this paragraph we clearly state the domain of applicability 
of our final results. Our derivation relies on several assumptions and approximations which are discussed in Sec.~\ref{approx}. First of all, 
the system has to be homogeneous at a scale which is smaller than or 
comparable to the characteristic inverse momentum and the inverse effective
mass, as required by the conditions 
(\ref{gradient-approx},\ref{quasi-particle-approx}). The system 
has to be weakly interacting so that the loop expansion is a legitimate 
approximation. This requires smallness of the coupling constants. Since 
we consider quasiparticles which are on the mass shell, the conditions 
(\ref{quasi-particle-approx}) and (\ref{quasi-particle-approx-2})
have to be fulfilled. The latter is easily satisfied if the bare
mass is much larger than the medium correction to the mass. A system 
of weakly interacting massive scalar fields which is close to 
equilibrium satisfies all of these conditions. 
 
In the future we plan to study further the multi-particle collision terms 
derived here. Although the integrals seem to be regular for massive
particles, they require very careful analysis, as the transport theory 
of dense gases or liquids is known to suffer serious problems at the level 
of multi-particle interactions, see {\it e.g.} \cite{Lif81}. The integral
represnting the process $3 \leftrightarrow 3$ is of particular concern
here as one particle can experience zero momentum transfer producing
a singularity of the respective propagator. 
  
We intend to extend the work of Refs. \cite{Jeon:1994if,Jeon:1995zm} on 
the role of multi-particle processes in transport phenomena. However,
we are going to start with the transport equations, which are derived 
here, and not with the Kubo formulas. We also plan to study how the system 
reaches chemical equilibrium due to particle production processes.
The first step in this direction is to compute the interaction rate  
\be
\label{rate}
\Gamma[n \ra m] \buildrel \rm def \over = 
\frac{1}{n} \int \frac{d^3p}{(2\pi)^3E_{p}} \: 
C^{(+)}[ p + p_1 + p_2 + \cdots p_{n-1} \ra q_1 + q_2 + \cdots + q_m]\;,
\ee
where $C^{(+)}[\cdots ]$ is the gain contribution to the collision 
term corresponding to the process 
$p + p_1 + p_2 + \cdots p_{n-1} \ra q_1 + q_2 + \cdots + q_m$.
Relative rates of the form $\Gamma[2 \ra 3]/\Gamma[2 \ra 2]$ or 
$\Gamma[3 \ra 3]/\Gamma[2 \ra 2]$ will give a measure of the importance of  
multi-particle processes.  

\acknowledgments

This project was initiated at the program `QCD and Gauge Theory Dynamics 
in the RHIC Era' organized by Kavli Institute for Theoretical Physics in 
Santa Barbara in April-June 2002. We are thus very grateful to the National 
Science Foundation for support under Grant No. PHY99-07949.



\begin{thebibliography}{99}

\bibitem{Han76}
J.P.~Hansen and I.R.~McDonald, {\it Theory of Simple Liquids}
(Academic Press, London, 1976).

\bibitem{Xiong:1992cu}
L.~Xiong and E.~V.~Shuryak,
Phys.\ Rev.\ C {\bf 49}, 2203 (1994)
[arXiv:hep-ph/9309333].

\bibitem{Geiger:1991nj}
K.~Geiger and B.~Muller,
Nucl.\ Phys.\ B {\bf 369}, 600 (1992).

\bibitem{Srivastava:1998vj}
D.~K.~Srivastava and K.~Geiger,
Nucl.\ Phys.\ A {\bf 647}, 136 (1999)
[arXiv:nucl-th/9806050].

\bibitem{Baier:2000sb}
R.~Baier, A.~H.~Mueller, D.~Schiff and D.~T.~Son,
Phys.\ Lett.\ B {\bf 502}, 51 (2001)
[arXiv:hep-ph/0009237].

\bibitem{Arnold:2002zm}
P.~Arnold, G.~D.~Moore and L.~G.~Yaffe,
JHEP {\bf 0301}, 030 (2003)
[arXiv:hep-ph/0209353].

\bibitem{Heinz:2004ik}
S.~M.~H.~Wong,
arXiv:hep-ph/0404222.

\bibitem{Xu:2004gw}
X.~M.~Xu, Y.~Sun, A.~Q.~Chen and L.~Zheng,
Nucl.\ Phys.\ A {\bf 744}, 347 (2004).
 
\bibitem{Jeon:1994if}
S.~Jeon,
Phys.\ Rev.\ D {\bf 52}, 3591 (1995)
[arXiv:hep-ph/9409250].

\bibitem{Jeon:1995zm}
S.~Jeon and L.~G.~Yaffe,
Phys.\ Rev.\ D {\bf 53}, 5799 (1996)
[arXiv:hep-ph/9512263].

\bibitem{Schwinger:1960qe}
J.~S.~Schwinger,
J.\ Math.\ Phys.\  {\bf 2}, 407 (1961).

\bibitem{Keldysh:ud}
L.~V.~Keldysh,
Zh.\ Eksp.\ Teor.\ Fiz.\  {\bf 47}, 1515 (1964)
[Sov.\ Phys.\ JETP {\bf 20}, 1018 (1965)].

\bibitem{Kad62} L.P. Kadanoff and G. Baym, 
{\it Quantum Statistical Mechanics} (Benjamin, New York, 1962).

\bibitem{Bez72} B. Bezzerides and D.F. Dubois, Ann. Phys. {\bf 70}, 10 (1972).

\bibitem{Li:1982gk}
S.~P.~Li and L.~D.~McLerran,
Nucl.\ Phys.\ B {\bf 214}, 417 (1983).

\bibitem{Danielewicz:kk}
P.~Danielewicz,
Annals Phys.\  {\bf 152}, 239 (1984).

\bibitem{Calzetta:1986cq}
E.~Calzetta and B.~L.~Hu,
Phys.\ Rev.\ D {\bf 37}, 2878 (1988).

\bibitem{Mrowczynski:1989bu}
St.~Mr\'owczy\'nski and P.~Danielewicz,
Nucl.\ Phys.\ B {\bf 342}, 345 (1990).

\bibitem{Botermans:1990qi}
W.~Botermans and R.~Malfliet,
Phys.\ Rept.\  {\bf 198}, 115 (1990).

\bibitem{Mrowczynski:1992hq}
St.~Mr\'owczy\'nski and U.~W.~Heinz,
Annals Phys.\  {\bf 229}, 1 (1994).

\bibitem{Henning:sm}
P.~A.~Henning,
Phys.\ Rept.\  {\bf 253}, 235 (1995).

\bibitem{Boyanovsky:1996xx}
D.~Boyanovsky, I.~D.~Lawrie and D.~S.~Lee,
Phys.\ Rev.\ D {\bf 54}, 4013 (1996)
[arXiv:hep-ph/9603217].

\bibitem{Boyanovsky:1999cy}
D.~Boyanovsky, H.~J.~de Vega and S.~Y.~Wang,
Phys.\ Rev.\ D {\bf 61}, 065006 (2000)
[arXiv:hep-ph/9909369].

\bibitem{Klevansky:1997wm}
S.~P.~Klevansky, A.~Ogura and J.~Hufner,
Annals Phys.\  {\bf 261}, 37 (1997)
[arXiv:hep-ph/9708263].

\bibitem{Mrowczynski:1997hy}
St.~Mr\'owczy\'nski,
Phys.\ Rev.\ D {\bf 56}, 2265 (1997)
[arXiv:hep-th/9702022].

\bibitem{Carrington:jt}
M.~E.~Carrington, D.~f.~Hou, A.~Hachkowski, D.~Pickering and J.~C.~Sowiak,
Phys.\ Rev.\ D {\bf 61}, 025011 (2000).

\bibitem{Carrington:2002bv}
M.~E.~Carrington, H.~Defu and R.~Kobes,
Phys.\ Rev.\ D {\bf 67}, 025021 (2003)
[arXiv:hep-ph/0207115].

\bibitem{Bez68}
B.~Bezzerides and P.~DuBois, Phys. Rev. {\bf 168}, 233 (1968).

\bibitem{Dan90}
P.~Danielewicz, Ann. Phys. {\bf 197}, 154 (1990).

\bibitem{Bozek:1997rv}
P.~Bo\.zek,
Phys.\ Rev.\ C {\bf 56}, 1452 (1997)
[arXiv:nucl-th/9704007].

\bibitem{Ivanov:1999tj}
Y.~B.~Ivanov, J.~Knoll and D.~N.~Voskresensky,
Nucl.\ Phys.\ A {\bf 672}, 313 (2000)
[arXiv:nucl-th/9905028].

\bibitem{Ivanov:2003wa}
Y.~B.~Ivanov, J.~Knoll and D.~N.~Voskresensky,
Phys.\ Atom.\ Nucl.\  {\bf 66}, 1902 (2003)
[arXiv:nucl-th/0303006].

\bibitem{Leupold:1999ga}
S.~Leupold,
Nucl.\ Phys.\ A {\bf 672}, 475 (2000)
[arXiv:nucl-th/9909080].

\bibitem{Leupold:xe}
S.~Leupold,
Acta Phys.\ Hung.\ New Ser.\ Heavy Ion Phys.\  {\bf 17}, 331 (2003).

\bibitem{Juchem:2003bi}
S.~Juchem, W.~Cassing and C.~Greiner,
Phys.\ Rev.\ D {\bf 69}, 025006 (2004)
[arXiv:hep-ph/0307353].

\bibitem{Juchem:2004cs}
S.~Juchem, W.~Cassing and C.~Greiner,
arXiv:nucl-th/0401046.

\bibitem{Gro80} S.R. deGroot, W.A. Van Leeuwen and Ch.G. van Weert,
{\it Relativistic Kinetic Theory} (North-Holland, Amsterdam, 1980).

\bibitem{Gelis:1997zv}
F.~Gelis,
Nucl.\ Phys.\ B {\bf 508}, 483 (1997)
[arXiv:hep-ph/9701410].

\bibitem{Weldon:jn}
H.~A.~Weldon,
Phys.\ Rev.\ D {\bf 28}, 2007 (1983).

\bibitem{Jeon:1998zj}
S.~Jeon and P.~J.~Ellis,
Phys.\ Rev.\ D {\bf 58}, 045013 (1998)
[arXiv:hep-ph/9802246].

\bibitem{Jeon2004}
S.~Jeon, private communication. 

\bibitem{Lif81}
E.M.~Lifshitz and L.P.~Pitaevskii, {\it Physical Kinetics}
(Pergamon Press, Oxford, 1981).

\end{thebibliography}
\end{document}